\documentclass[fleqn,usenatbib]{mnras}

\usepackage{newtxtext,newtxmath}
\usepackage[T1]{fontenc}
\usepackage{ae,aecompl}

\usepackage{graphicx}	% Including figure files
\usepackage{amsmath}	% Advanced maths commands
\usepackage{ulem}
\usepackage{booktabs}
\usepackage{soul}	% Pour barrer du texte
\usepackage{xcolor}
\newcommand{\beq}{\begin{equation}}
\newcommand{\eeq}{\end{equation}}
\newcommand{\bea}{\begin{eqnarray}}
\newcommand{\eea}{\end{eqnarray}}
\newcommand{\nn}{\nonumber}

\newcommand{\erre}{{\cal R}}
\newcommand{\Dif}{{\cal D}}
\usepackage{color}
\usepackage{subfigure}
\usepackage{multirow}
\usepackage{float}
\usepackage{nameref}
\usepackage{natbib}
\usepackage{soul}	
\usepackage{xcolor}

\graphicspath{{./figures/}}

\title[]{High energy cosmic rays and gamma rays from star clusters: the case of Cygnus OB2}

\author[]{
	Pasquale Blasi$^{1,2}$\thanks{E-mail: pasquale.blasi@gssi.it} \&
	Giovanni Morlino $^{3}$\thanks{E-mail: giovanni.morlino@inaf.it},
	\\
	$^{1}$ Gran Sasso Science Institute, Viale F. Crispi 7 - 67100 L' Aquila, Italy\\
	$^{2}$ INFN-Laboratori Nazionali del Gran Sasso, Via G. Acitelli 22, Assergi (AQ), Italy \\
	$^{3}$ INAF/Osservatorio Astrofisico di Arcetri, Largo E. Fermi, 5 - 50125 Firenze, Italy
}

\date{}

\pubyear{2021}

\begin{document}

\label{firstpage}
\pagerange{\pageref{firstpage}--\pageref{lastpage}}
\maketitle

% Abstract of the paper
\begin{abstract}
We investigate the acceleration of cosmic rays at the termination shock that results from the interaction of the collective wind of star clusters with the surrounding interstellar medium. The solution of the transport equation of accelerated particles in the wind-excavated cavity, including energy losses due to CR interactions with neutral gas in the bubble, shows several interesting properties that are discussed in detail. The issue of the maximum energy of the accelerated particles is discussed with special care, because of its implications for the origin of Galactic cosmic rays. Gamma ray emission is produced in the cavity due to inelastic pp scattering, while accelerated particles are advected downstream of the termination shock and diffuse at the same time. Both the spectrum and the morphology of such emission are discussed, with a comparison of our results with the observations of gamma ray emission from the Cygnus OB2 region.
\end{abstract}

% Select between one and six entries from the list of approved keywords.
% Don't make up new ones.

%%%%%%%%%%%%%%%%%%%%%%%%%%%%%%%%%%%%%%%%%%%%%%%%%%

%%%%%%%%%%%%%%%%% BODY OF PAPER %%%%%%%%%%%%%%%%%%

%%% INTRODUCTION %%%
\section{Introduction}
\label{sec:introduction}

Recent observations of gamma ray emission from some stellar clusters, such as Westerlund 1 \citep{Abramowski_Wd1:2012}, Westerlund 2 \citep{Yang+2018}, the Cygnus cocoon \citep{Ackermann:2011p3159,Aharonian+2019NatAs}, NGC 3603 \citep{Saha_NGC3603:2020}, BDS2003 \citep{HAWC:2021}, W40 \citep{Sun_W40:2020} and 30 Doradus in the LMC \citep{HESS-30Dor:2015}, have revived the interest in these astronomical objects as potential sites of particle acceleration \citep{Gupta+2018,Bykov+2020,2021MNRASMorlino,Bykov-Kalyashova:2022, Vieu+2022}, a topic first discussed by \cite{Cesarsky-Montmerle:1983,webb1985}. In particular, the detection of gamma rays with energy up to $\sim 100$ TeV with HAWC \cite[]{2021NatureHAWC} and up to 1.4 PeV with LHAASO from the Cygnus Cocoon region \citep[]{LHAASO-2021Nature} has boosted the investigation of these sources as accelerators of cosmic rays (CRs) up to the knee.  

This scenario has attracted even more attention in recent years given the challenges faced by the standard paradigm for the origin of CRs, based on supernova remnants (SNRs) as chief sources of CRs. Particle acceleration at the forward shock \cite[see][for a review]{blandford} in typical type Ia and standard core collapse supernovae faces daunting difficulties in reaching energies as high as the knee, both with resonant streaming instability \cite[]{Lagage1,Lagage2} and with the faster growing non-resonant instability \cite[]{bell2004}, as also discussed by \cite{Schure-Bell:2013} and \cite{pierre}. A possible exception to this conclusion could be powerful ($\gtrsim 5 \times 10^{51}$ erg) and rare ($\sim 1/10^{4}$ years) core collapse SNRs, with relatively small ejecta mass (few solar masses), for which the maximum energy can reach PeV energies \cite[]{pierre,pierre2021}. However, given the low rate of these supernova events, the chances of detecting gamma rays from one of these sources in our Galaxy appear to be exceedingly dim. 

Given these difficulties, it is only natural to look for alternative accelerators of Galactic CRs, such as star clusters and the shocks developed in their winds \cite[]{2021MNRASMorlino}. Moreover CR acceleration in winds of star clusters would also help addressing the well known problem of the anomalous $^{22}$Ne/$^{20}$Ne abundance in CRs \citep{Gupta+2020,Kalyashova-Bykov:2021,Tatischeff+2021}. In fact this ratio is a factor $\sim 5$ larger in CRs than in the solar environment \citep{Binns+:2006}, a result that is difficult to accommodate in the framework of particle acceleration at SNR shocks alone \citep{Prantzos2012}.

The recent measurement of the spectrum and spatial distribution of the gamma ray emission from the Cygnus cocoon \cite[]{2021NatureHAWC} represents a unique opportunity to test our understanding of particle acceleration in the environment around star clusters. The theoretical description of the process of particle energization and transport in the cavity excavated by the wind launched by a star cluster was recently outlined by \cite{2021MNRASMorlino}, where the authors focused on the acceleration at the wind termination shock, discussing the maximum momentum of the accelerated particles and its dependence upon the diffusion coefficient in the region of the termination shock. Here we use the same theoretical framework but we solve numerically the transport equation for protons with the inclusion of energy losses. Such phenomenon, dominated by inelastic collisions, is also responsible for the bulk of the gamma ray emission from star clusters, while it has no appreciable effect on the maximum energy of the accelerated particles, since the acceleration process is typically very fast. On the other hand, the spectrum of CR protons propagating downstream  of the termination shock and the spectrum of particles escaping the cavity excavated by the wind can be considerably affected by such losses, at least for massive clusters. This effect shapes the spectrum of CR protons released into the ISM, at least for particles with energies such that their transport in the cavity is dominated by advection. We specialize these general predictions to the case of the Cygnus cocoon, for which detailed spectral and morphological information are now available.

The article is organised as follows: in \S \ref{sec:bubble} we briefly summarize the general properties of the wind blown cavity and the role of cooling and clumpiness in the distribution of cold gas. In \S \ref{sec:Pmax} we discuss the main considerations that enter the calculation of the maximum energy of accelerated particles at the termination shock. In \S~\ref{sec:theory} we describe the numerical solution of the transport equation of non-thermal particles in the cavity and the associated diffusive particle acceleration at the termination shock. In \S \ref{sec:results} we describe our results in terms of spectrum of accelerated particles and gamma ray emission. We specialize our findings to the description of the spectrum and spatial morphology of the gamma ray emission from the Cygnus cocoon. In \S \ref{sec:concl} we outline our conclusions.
 
%%% SECTION %%%
\section{The wind blown bubble}
\label{sec:bubble}

\begin{figure}
\centering
\includegraphics[width=.45\textwidth]{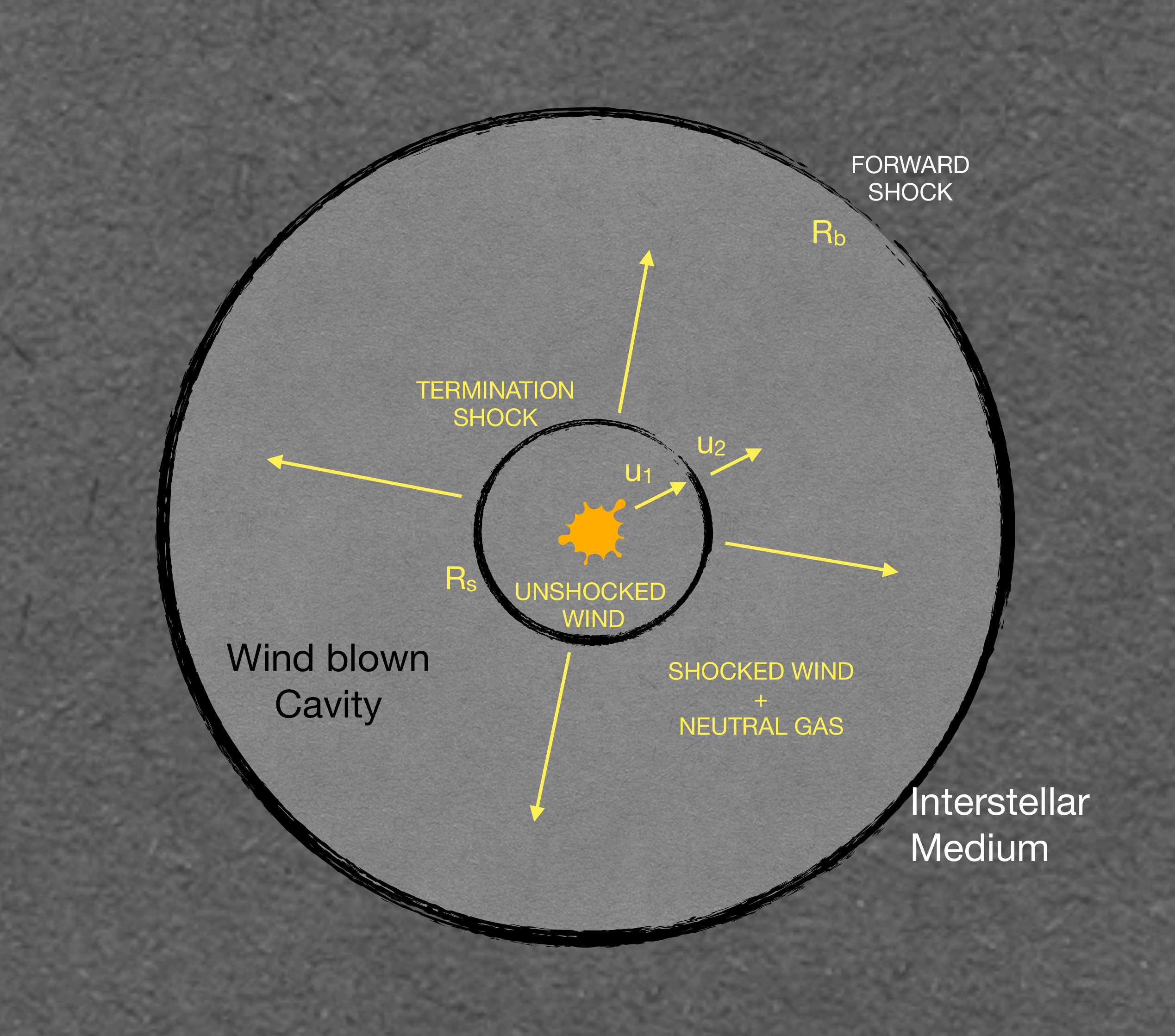}
\caption{Schematic structure of a wind blown bubble excavated by a star cluster into the ISM: $R_{s}$ is the position of the termination shock, $R_{\rm b}$ is the radius of the forward shock.}
\label{fig:geometry}
\end{figure}

In Figure~\ref{fig:geometry} we show a schematic view of the cavity blown by the collective wind of the stars located in the central region. We explicitly assume here to be dealing with a compact star cluster, namely a cluster in which the termination shock is located well outside the region where the stars are concentrated. 

Immediately outside the stellar cluster, the winds of the individual objects merge into a collective wind, with a velocity $v_{w}$. The wind density is obtained from mass conservation:
\beq
\rho_w(r) = \frac{\dot M}{4\pi r^{2} v_{w}},~~~ r > R_{c},
\eeq
where $R_{c}$ is the radius of the core where the stars are concentrated, and $\dot M$ is the rate of mass loss due to the collective wind. The impact of the supersonic wind with the ISM, assumed here to have a constant density $\rho_{0}$, produces a forward shock at position $R_{\rm b}$, while the shocked wind is bound by a termination shock, at a location $R_{s}$. The shocked ISM and the shocked wind are separated by a contact discontinuity (not shown in Figure~\ref{fig:geometry}), very close to the forward shock. The region between the contact discontinuity and the forward shock should contain dense cold interstellar gas, plowed away during the expansion of the cavity. However, several instabilities are expected to spread this gas inside the cavity while the bubble is being blown. Hence clouds of dense molecular gas and regions of dense atomic gas are expected to fill, more or less homogeneously, the cavity. We will refer to the density of this gas as $\rho$ (we will use the symbol $n$ to indicate the number density in the bubble). One can easily check that for typical values of the parameters, the density contributed by the wind is completely negligible, especially in terms of gamma ray production. 

Since the typical cooling timescale of the shocked ISM is only $\sim 10^4$ yr, while the cooling time for the shocked wind is several $10^7$ yr \citep{Koo-McKee:1992a, Koo-McKee:1992b}, we can safely assume that the wind-blown bubble evolves quasi-adiabatically. Following \cite{Weaver+1977} and \cite{Gupta+2018}, \cite{2021MNRASMorlino} provided some useful approximations for the position of the forward shock (FS) and the termination shock (TS), that we use here. The position of the FS is at
\beq \label{eq:Rbubble}
 R_b(t) = 139  ~ \rho_{10}^{-1/5} \dot{M}_{-4}^{1/5}v_{8}^{2/5} t_{10}^{3/5} ~\rm pc,
\eeq
where $\rho_{10}$ is the ISM density in the region around the star cluster in units of 10 protons per cm$^{3}$, $v_{8}=v_{w}/(1000~\rm km\, s^{-1})$, $\dot M_{-4}=\dot M/(10^{-4}\rm M_{\odot}\,yr^{-1})$ and $t_{10}$ is the dynamical time in units of 10 million years. The wind luminosity is then $L_{w}=\frac{1}{2}\dot M v_{w}^{2}$. 
The termination shock is located at 
\beq \label{eq:RTS}
 R_{s}= 24.3 ~ \dot{M}_{-4}^{3/10} v_{8}^{1/10} \rho_{10}^{-3/10} t_{10}^{2/5} ~\rm pc \,.
\eeq

A more accurate calculation \citep{Weaver+1977} shows that the results above are accurate within $\lesssim 10\%$. We stress again that the speed of the TS in the laboratory frame is very low, so that the entire bubble structure evolves slowly and can be considered as stationary to first approximation. It is worth stressing that the formation of a collective wind occurs only for compact clusters that have a typical cluster size $R_c \ll R_{s}$ \cite[see, e.g.][]{Gupta+2020}.
 
\subsection{Cooling effects and clumpiness} \label{sec:clumps}
In the model described above, the volume averaged density and temperature of the hot shocked wind can be estimated as
\beq  \label{eq:n_bubble}
  n_{\rm b} = \frac{\dot{M} \, t_{\rm age}}{4\pi /3 \,R_{\rm b}^3}
  = 3.6 \times 10^{-3} \rho_{10}^{3/5} \, \dot{M}_{-4}^{2/5} \, v_{8}^{-6/5} \, t_{10}^{-4/5} \, \rm cm^{-3} \,
\eeq
and
\beq  \label{eq:T_bubble}
  T_{b} = \frac{P}{n_b k_{\rm B}} 
  \simeq 10^{7}\, L_{w,38}^{2/5}\, n_{10}^{3/5}\, n_{b,-2}^{-1} t_{10}^{-4/5}\, \rm K \,,
\eeq
respectively, where we introduced $n_{b,-2}=n_b/10^{-2}\rm cm^{-3}$. On the other hand, cooling, that we neglected in the estimates above, leads to a reduction of the temperature and therefore a smaller size of the bubble. \cite{Gupta+2016} retained the effect of cooling and accounted for the radiation pressure from the stars, using 1D hydro-dynamical simulations. In this way, they predict a bubble size that is smaller by $\sim 30\%$ at an age of a few Myr and a temperature roughly one order of magnitude smaller than the one estimated in Eq.~\eqref{eq:T_bubble}. These effects appear rather mild in terms of the global structure of the bubble and the associated high energy phenomenology. However, the simulations of \cite{Gupta+2016}, being 1D in space, do not account for the possible presence of clumps in the bubble that may enhance the effect of cooling. 

The presence of dense clumps in the bubble is especially important for the problem discussed here, in that it may affect the strength and morphology of the gamma ray signal. As we discuss in \S~\ref{sec:results}, the present gamma ray observations show that the average gas density in the bubble, as a target for hadronic interactions, is required to be much larger than the mean wind density estimated above. This gas is likely distributed in the form of dense clumps and here we discuss the implications of such a clumpy distribution. The clumps may either originate from a pre-existing molecular complex in the region or can result from the fragmentation of the compressed ISM shell, due to the onset of hydro-dynamical instabilities.

In the following we limit ourselves to discuss under which conditions the presence of clumps may change significantly the bubble structure with respect to the adiabatic model developed by \cite{Weaver+1977} and adopted here.

The hot bubble is mainly affected through two different processes: radiative losses and thermal conduction with cold material. While the former reduces the pressure in the bubble, hence reducing its size, the latter produces evaporation of the cloud material, hence decreases the bubble temperature but does not reduce its pressure. The radiative cooling timescale is
\beq    \label{eq:tau_rad}
  \tau_{\rm rad} = \frac{3/2 \, k_{\rm B} T_{\rm b} \, n_b}{\Lambda \, n_b^2} = 6 \times 10^6 \, n_{\rm b,-2}^{-1} \, T_6^{1.7} \, \rm yr,    
\eeq
where $\Lambda$ is the cooling rate which in the temperature interval between $10^5$ and $10^7$\,K is well approximated by $\Lambda= 1.1 \times 10^{-22} T_6^{-0.7}\, \rm erg\, cm^{3}\, s^{-1}$ \citep{Draine_book:2011}. Hence, radiative losses alone could be neglected for the first few Myr, although, as we discuss below, accounting for thermal conduction may change this conclusion.

Let us assume that clumps have typical density $n_{\rm cl}$ and radius $R_{\rm cl}$. The volume filling factor of the clumps is given by $\chi_{\rm cl} \simeq \bar{n}/n_{\rm cl}$, where $\bar{n}$ is the average bubble density as will be inferred from gamma-ray emission, while the number of clumps is $N_{\rm cl} = \chi_{\rm cl} (R_{\rm b}/R_{\rm cl})^3$. 
The ideal value of the rate of thermal conduction per unit surface, which also maximizes the value of this quantity, is $q = -\kappa_s \nabla T$ with $\kappa_s \simeq 6.1 \times 10^{-7} \, T^{5/2} \, \rm erg \, s^{-1} \, K^{-1}\, cm^{-1}$ \citep{spitzerBook}. We further approximate the large scale temperature gradient in the bubble as $\nabla T \simeq T_b/R_b$ while the gradient close to a clump is $\nabla T\simeq T_b/R_{\rm cl}$. To estimate the total conduction timescale, we consider the contact surface with both the cold ISM shell ($\Sigma_{\rm sh}= 4\pi R_b^2$) and the clumps ($\Sigma_{\rm cl} = 4\pi N_{\rm cl} R_{\rm cl}^2$). Hence, the conduction timescale can be estimated as 
\beq    \label{eq:tau_cond}
  \tau_{\rm cond} = \frac{3/2 k_B T_{\rm b} n_{\rm b} V_{\rm b}}{q_{\rm sh}  \Sigma_{\rm sh} + q_{\rm cl}  \Sigma_{\rm cl}}
  = 3.5 \times 10^6 \, \frac{n_{\rm b,-2} \, R_{\rm b,2}^2 \, T_{\rm b, 6}^{-5/2}}{1+ \chi_{\rm cl} \left( R_{\rm b}/R_{\rm cl}\right)^2} \, \rm yr \,.
\eeq
The cooling timescale is reduced by the presence of clumps through the term $\chi_{\rm cl} (R_b/R_{\rm cl})^2$ which is smaller than unity if $n_{\rm cl} > \bar{n}(R_b/R_{\rm cl})^2$.
For the case of Cygnus cocoon analyzed below we will use $\bar{n} \approx 10\div 20 \, \rm cm^{-3}$, while $R_{\rm b}\simeq 100\,\rm pc$, hence the above condition will be satisfied provided $n_{\rm cl} \gtrsim 10^5 (R_{\rm cl}/{\rm pc})^{-2} \rm \, cm^{-3}$. 
Now, the equilibrium temperature in the bubble can be estimated equating the age of the system with the total cooling timescale, i.e. $t_{\rm age}^{-1} = \tau_{\rm rad}^{-1} + \tau_{\rm cond}^{-1}$. Using $n_{\rm cl}=10^3$ and $R_{\rm cl}=3$\,pc we get $T_{\rm b} \approx 10^6$\,K, in reasonable agreement with the result from numerical simulations obtained by \cite{Gupta+2016}.

One should also notice that the density of the hot plasma in the bubble will be enhanced by evaporation from the cold material. The classical evaporation rate of a spherical clump embedded in a hot fully ionized plasma is given by $\dot{M}_{\rm ev} = 16 \pi \, \mu \, \kappa_s \, R_{\rm cl}/(25 k_{\rm B})$, where $\mu$ is the mean mass per particle \citep{Cowie-McKee:1977}. The same expression applies for the entire shell, hence the full evaporation rate is
\begin{eqnarray} \label{eq:Mdot_ev}
  \dot{M}_{\rm ev} = \frac{16 \pi \, \mu \kappa_s}{25 k_{\rm B}} (R_b + N_{\rm cl} R_{\rm cl}) \hspace{3cm} \nonumber \\
  \simeq  4.4 \times 10^{-5} \,T_{\rm b,6}^{5/2} \frac{R_{b}}{100\,{\rm pc}} \left(1 + \chi_{\rm cl} \frac{R_b^2}{R_{\rm cl}^2} \right)
  \rm M_{\odot} \,yr^{-1} \,. 
\end{eqnarray}  
Similarly to Eq.\eqref{eq:tau_cond}, Eq.\eqref{eq:Mdot_ev} shows that the evaporated mass is dominated by clumps if $\chi_{\rm cl} > (R_{\rm cl}/R_b)^2$. If we assume $n_{\rm cl}\approx 10^3\, \rm cm^{-3}$ and $R_{\rm cl}\approx 3\,\rm pc$ the mass loss rate from evaporation is of the same order than the one from the stellar winds ($\dot{M}_{\rm wind} \sim 10^{-4}\,\rm  M_{\odot}\, yr^{-1}$, see Table~\ref{tab:Models}).
If $\dot{M}_{\rm ev}$ becomes comparable with or larger than the mass loss rate due to stellar winds, then $\tau_{\rm rad}$ decreases while $\tau_{\rm cond}$ increases, thereby leaving the equilibrium temperature almost unchanged. 
In conclusion, it is reasonable to assume that clumps have a rather marginal effect on the structure of the bubble, compared with the pure effect of cooling, resulting in a bubble modification similar to the one obtained by \cite{Gupta+2016}, namely with a size reduced by $\sim 30\%$. If this is the case, then our results for Cygnus OB2 are not expected to be affected in a significant way, as we discuss in \S~\ref{ssec:gamma}. However, a more detailed assessment of this issue requires one to account for the time evolution of the clump structure together with the wind-bubble, something that will be considered in a forthcoming piece of work.

\subsection{Particle propagation}
Given the assumption of approximately adiabatic evolution of the bubble, the gas density downstream of the TS is roughly constant, hence, from mass conservation it follows that the plasma velocity drops as $1/r^2$. These scaling relations, previously described in detail by \cite{2021MNRASMorlino}, are used below in solving the transport equation for non-thermal particles. 

Typically the core of a massive stellar cluster can contain up to $\sim 100-1000$ stars whose winds interact strongly leading to partial dissipation of kinetic energy of the winds, which may result in generation of turbulent magnetic field in the free expanding wind \citep{Badmaev+2022}. This implies that the collective wind outside the core is not expected to have a coherent, spiral-like structure, as one might be tempted to assume for the wind of an individual star. 

If a fraction $\eta_B$ of the kinetic energy of the wind is transformed to magnetic energy, at the termination shock one may expect a turbulent magnetic field of order 
\beq \label{eq:B_TS}
B(R_{s})=7.4 \; \eta_{B}^{1/2} \dot M_{-4}^{1/5}v_{8}^{2/5} \rho_{10}^{3/10} t_{10}^{-2/5}~ \,{\rm \mu G}.
\eeq
This dissipation of kinetic energy into magnetic energy likely results in turbulence with a typical scale $L_c$ that is expected to be of order the size of the star cluster, $L_c\sim R_c\sim 1\div 2$ pc. 

Assuming that the turbulence follows a Kraichnan cascade, the diffusion coefficient upstream of the TS can be estimated as 
\begin{flalign}  \label{eq:DKra}
 D(E)\approx\frac{1}{3} r_{L}(p) v \left(\frac{r_{L}(p)}{L_{c}}\right)^{-1/2} = 
 1.1\times 10^{25} 
 \left( \frac{L_c}{1\rm pc}\right)^{1/2} \hspace{0.9cm} \nn \\
  \eta_{B}^{-1/4} \dot M_{-4}^{-1/10} v_{8}^{-1/5} \rho_{10}^{-3/20} t_{10}^{1/5} E_{\rm GeV}^{1/2}~\rm cm^2~s^{-1},
\end{flalign}
where $r_{L}(p)=pc/e B(r)$ is the Larmor radius of particles of momentum $p$ in the magnetic field $B(r)$. Here we introduced the energy $E=pc$ of relativistic particles.

Other types of turbulent spectra were discussed by \cite{2021MNRASMorlino}. While the estimated maximum energy does not change dramatically with different choices of the turbulent cascade, the shape of the spectrum of accelerated particles is sensibly affected by such choice. For instance, for a Kolmogorov spectrum, the spectrum of accelerated particles smoothly softens towards high energies and this results in an effective maximum energy that is inadequate to describe gamma ray spectra that extend to the $\gtrsim 100$ TeV energy range \cite[see][for an extensive discussion of this issue]{Menchiari2023}.

The effectiveness of the process of dissipation of wind kinetic energy to magnetic energy is poorly constrained and it is worth exploring the possibility that CR induced instabilities may be efficient in producing magnetic perturbations in the wind region. If the field carried by the wind is particularly low (see below for a more quantitative estimate), the non-resonant hybrid instability \citep{Bell:2004p737} may excite magnetic perturbations on spatial scales that, at saturation of the instability, are expected to be comparable with the Larmor radius of the particles dominating the driving current at a given location. In this case, the maximum value of the pressure in the form of amplified magnetic field at the shock reads:

\beq \label{eq:B_non-res}
  \frac{\delta B_1^2}{4 \pi} = \frac{v_s}{c}\, \rho_1 v_s^2 \, \frac{\xi_{\rm CR}}{\Lambda},
\eeq
where we assumed a spectrum of accelerated particles $\propto p^{-4}$ and we introduced $\xi_{\rm CR}$ as the fraction of ram pressure $\rho_1 u_w^2$ that gets channelled into CR pressure, and $\Lambda=\ln\left(\frac{p_{max}}{m_p c}\right)\sim 10$ (see also Eq. \ref{eq:nCR}). This condition can be achieved if there is enough time for the instability to grow for a sufficient number of e-folds (typically $\sim 5-10$ e-folds would suffice), as we discuss below in \S~\ref{sec:Pmax}. For values of the parameters that are typical of a massive star cluster, the saturated  magnetic field would be
\beq \label{eq:B_non-res2}
  \delta B_1 \approx 0.043 \; \frac{\xi_{\rm CR}/0.05}{\Lambda/10}\, \dot{M}_{-4}^{1/5} \, v_8^{9/10} \, \rho_{10}^{3/10} \, t_{10}^{-2/5} \;\rm \mu G \,.
\eeq
This self-generated turbulence would be such that, for a spectrum of accelerated particles close to $\sim E^{-2}$, the corresponding diffusion coefficient would be Bohm-like. Spectra steeper than $E^{-2}$ would lead to a lack of power on large scales and a more pronounced energy dependence than linear in the diffusion coefficient. If the Bohm condition could be achieved, the corresponding coefficient would read:
\beq \label{eq:D_selfgen}
  D(E)\approx\frac{r_{L} v}{3}  =  
  7.7 \times 10^{22} \, E_{\rm GeV} \dot{M}_{-4}^{-\frac{1}{5}} \, v_8^{-\frac{9}{10}} \rho_{10}^{-\frac{3}{10}} \, t_{10}^{\frac{2}{5}} \; \rm cm^2\, s^{-1}.
\eeq
The excitation of the non resonant instability occurs only if the pre-existing magnetic field is smaller than the value in Equation~\eqref{eq:B_non-res2}. Moreover, as discussed below, the saturation level in Equation~\eqref{eq:B_non-res2} is reached only several e-folds are allowed during one advection time upstream. 

Independent of the choice of the diffusion coefficient, downstream of the termination shock, we assume that the magnetic field is only compressed by the standard factor $\sqrt{(2 \mathcal{R}^2+1)/3}$, that for a strong shock (compression factor $\mathcal{R}=4$) becomes $\sqrt{11}$. For the Kraichnan case, $D_{2}\approx 0.55 D_{1}$. Clearly the downstream diffusion coefficient can be smaller than this estimate suggests, if other processes, perhaps of hydro-magnetic origin, such as the Richtmyer-Meshkov instability \citep{Giacalone:2007p962}, lead to enhanced turbulence behind the shock. We will discuss some implications of this scenario below. 

The functional shape of the diffusion coefficient as in Equation~\eqref{eq:DKra} is expected to hold up to energies for which the Larmor radius equals the coherence scale $L_c$. At larger energies the standard $D(E)\propto E^2$ appears, as can be found both analytically and using simulations of test particle transport in different types of synthetic turbulence \cite[see for instance][and references therein]{Subedi2017,Dundovic2020}.

\section{Maximum momentum}
\label{sec:Pmax}

An estimate of the maximum energy that can be achieved at the TS through DSA can be easily obtained even without a formal solution of the transport equation, although, as discussed by \cite{2021MNRASMorlino}, special care is needed in interpreting the physical meaning of such maximum momentum: due to the combination of spherical symmetry of the problem and different energy dependence of the diffusion coefficient, the spectrum of accelerated particles is characterized by a pronounced cutoff at the maximum momentum in the case of Bohm diffusion, while a milder energy dependence in $D(E)$ results in a gradual roll-off, more similar to a spectral steepening that starts at $p\ll p_{\max}$. The case of a Kraichnan turbulence is sort of intermediate between Bohm and Kolmogorov and, as we discuss below, provides the best description of the available observations.  

The maximum momentum is defined by the most stringent among the following conditions:
\begin{enumerate}
    \item[$1)$] the diffusion length upstream must be smaller that the radius of the termination shock: for the Kraichnan case this condition reads
    \beq \label{eq:Emax_cond1}
        p_{\rm max}^{(1)}= 4\times 10^5 \left( \frac{L_c}{1\rm pc}\right)^{-1} \eta_{B}^{1/2} \dot M_{-4}^{4/5} v_{8}^{13/5} \rho_{10}^{-3/10} t_{10}^{2/5} \; \rm GeV/c.
    \eeq
    If the non-resonant streaming instability is excited and can saturate to its nominal value (Bohm diffusion in Equation~\eqref{eq:D_selfgen}), this condition becomes
    \beq
        p_{\rm max}^{(1)}=9.8\times 10^3 \dot M_{-4}^{1/2} v_{8}^{2} \; \rm GeV/c.
    \eeq
    \item[$2)$] The diffusion length downstream must not exceed the size of the downstream region, which implies, for the Kraichnan case:
    \beq \label{eq:Emax_cond2}
        p_{\rm max}^{(2)}= 2.7\times 10^5 \left( \frac{L_c}{1\rm pc}\right)^{-1} \eta_{B}^{1/2} \dot M_{-4}^{3/5} v_{8}^{16/5} \rho_{10}^{-1/10} t_{10}^{4/5} \; \rm GeV/c.
    \eeq
    For the case of Bohm self-generated diffusion, the condition reads:
    \beq
        p_{\rm max}^{(2)}= 4.7\times 10^4 \dot M_{-4}^{2/5} v_{8}^{23/10} \rho_{10}^{1/10} t_{10}^{1/5} \; \rm GeV/c.
    \eeq
    \item[$3)$] For the Kraichnan case the scattering should occur in the inertial range of the turbulence, namely the Larmor radius should not exceed the coherence scale $L_c$:
    \beq \label{eq:Emax_cond3}
        p_{\rm max}^{(3)}=6.8\times 10^5 \left( \frac{L_c}{1\rm pc}\right)^{-1} \eta_{B}^{1/2} \dot M_{-4}^{1/5} v_{8}^{2/5} \rho_{10}^{3/10} t_{10}^{-2/5} \; \rm GeV/c.
    \eeq
\end{enumerate}
For the case of self-generated turbulence, this latter condition is replaced by the requirement that the non-resonant instability has enough time to grow. If $v_A$ is the Alfv\'en speed in the pre-existing magnetic field $B_0$, the growth rate of the instability can be estimated as $\gamma_W=k_m v_A$, where $k_m$ is the wavenumber where the instability grows the fastest, namely $k_m B_0 = \frac{4\pi}{c} \, e \, n_{\rm CR} \,v_w$, and the density of CR particles at the shock is estimated as
\beq
n_{CR}(>E) = \frac{\xi_{CR}\, \rho_w(R_s) \, v_w^2}{\Lambda \, E}, 
\label{eq:nCR}
\eeq
and $\Lambda\sim \ln(E_{\max}/E_{\min})\sim 10$. It is worth stressing that $\gamma_W$ does not depend upon the pre-existing magnetic field $B_0$, and can be written as:
\beq
\gamma_W=5\times 10^{-9} \dot M_{-4}^{1/5} v_{8}^{12/5} \rho_{10}^{3/10} t_{10}^{-2/5} E_{\rm GeV}^{-1} \; \rm s^{-1}. 
\eeq
If we introduce the advection time upstream, $\tau_{\rm adv}=R_s/v_w$, the condition that the non-resonant instability grows is that $\gamma_W \tau_{\rm adv}\sim \zeta$, with $\zeta\sim 5-10$. This condition results in an upper limit on the energy of the particles that can possibly be accelerated at the termination shock
\beq \label{eq:Emax_non-res}
  p_{\max} = 3.7\times 10^3 \, \zeta^{-1}\, v_8^{3/2} \, \dot{M}_{-4}^{1/2} \; \rm GeV/c.
\eeq
One can easily see that even for the unrealistic value $\zeta\sim 1$, this condition limits the maximum energy of the accelerated particles to be exceedingly small and in any case too low to account for the high energy gamma ray emission of the Cygnus cocoon. This is due to the fact that the non-resonant instability grows too slowly in the upstream plasma to allow for turbulence to grow and scatter particles. Based on this finding, we conclude that the assumption of Bohm diffusion is, in this context, not justified and in the rest of the discussion below we focus on the Kraichnan case. 

The actual value of the maximum momentum is determined by the most stringent condition among those listed above, which depends upon the values of the parameters (mass loss rate, wind speed, density of the ISM, age of the star cluster, efficiency of conversion to magnetic turbulence and coherence scale of the turbulence). The strongest dependence is the one on the wind speed. 
\begin{figure}
\centering
\includegraphics[width=.45\textwidth]{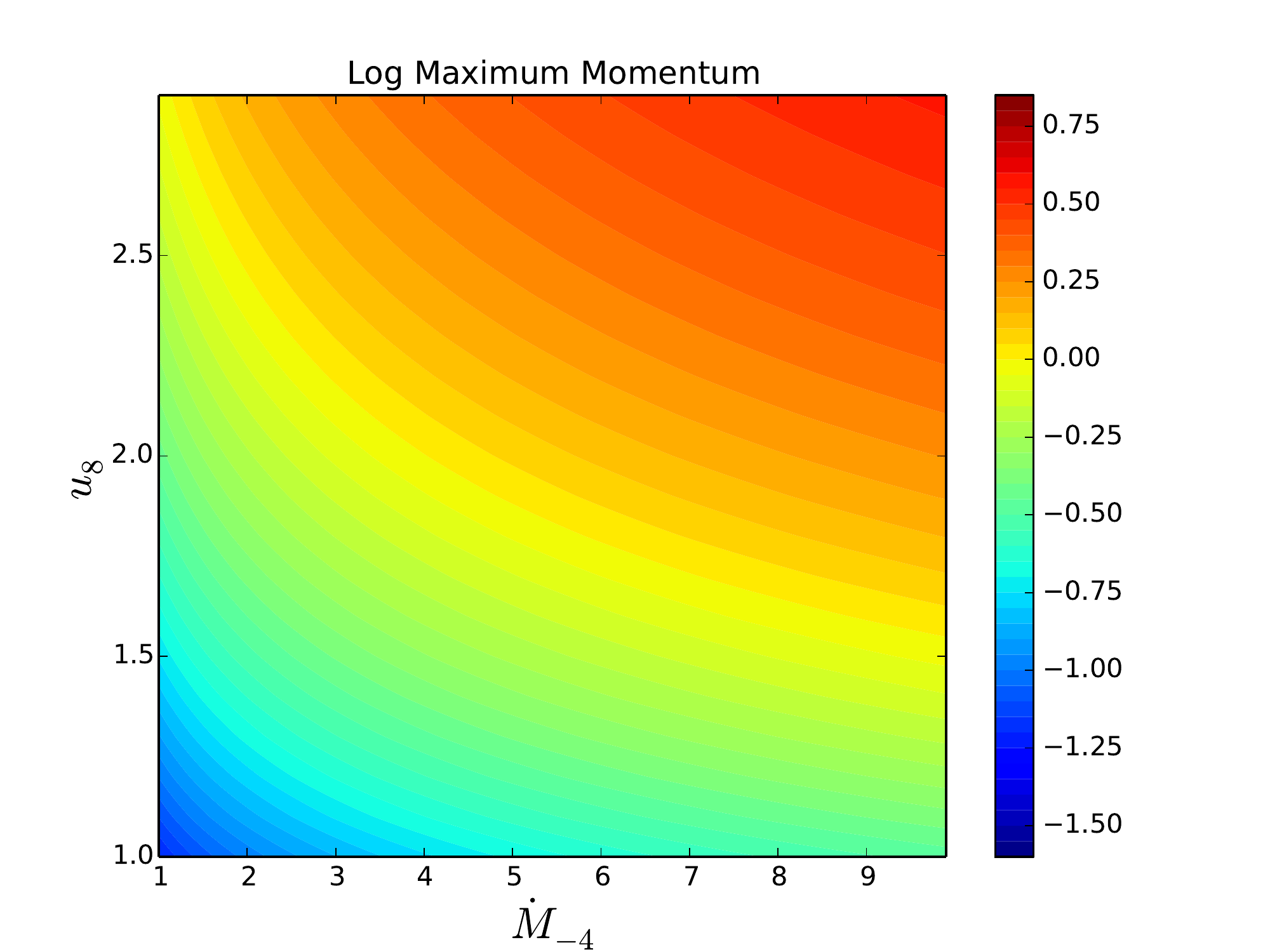}
\caption{Contour plot of the $\log(c p_{\rm max}/\rm PeV)$ as a function of the rate of mass loss and the wind speed.}
\label{fig:Pmax}
\end{figure}
For the case of Kraichnan turbulence the dependence of the maximum momentum on the mass loss rate and the wind speed is illustrated in the contour plot in Figure~\ref{fig:Pmax}, where we show $\log(p_{\rm max}c/\rm PeV)$ for parameters that are thought to be appropriate for the Cygnus cocoon (age of 3 million years, density of the ISM outside the cavity of $\sim 10 ~\rm cm^{-3}$ and coherence scale of Kraichnan turbulence chosen as $L_c=1$ pc). One can see that for the maximum momentum to fall in the range around 1 PeV ($\log( p_{\rm max}/\rm PeV)\simeq 0$), either very fast winds or large rates of mass loss are required. We will see below that even these conditions may not be sufficient to make a typical star cluster a PeVatron. 

In the discussion above and in all of this article we have not considered the spatial dependence of the diffusion coefficient. It may be useful to comment on the possible effects of such an assumption, especially in the direction of considering effects that may reduce the diffusion coefficient and lead to potentially larger values of the maximum momentum of accelerated particles. Let us start from the upstream region and assume that the diffusion coefficient upstream of the termination shock depends on $r$, in the direction of becoming smaller moving inward. One way (certainly not unique) to do so is to assume that the same fraction of kinetic energy of the wind at distance $r$ is converted to magnetic energy. It is hard to imagine a stronger radial dependence, in that it would easily lead to more than $10\%$ conversion efficiency at small radii, with huge dynamical implications for the wind. If $\eta_B$ is independent of $r$, the resulting magnetic field scales as $(r/R_s)^{-1/2}$ (see Eq. \ref{eq:B_TS}). For a Kraichnan turbulence spectrum, this would translate to a diffusion coefficient $D(E,r)=D(E) (r/R_s)^{1/4}$, where $D(E)$ is given in Eq. \ref{eq:DKra}. Clearly the dependence on $r$ is very weak: for a given energy, reducing the diffusion coefficient by one order of magnitude would require to reach radii $r\sim 10^{-4}R_s$, well inside the core of the star cluster, where none of this description would apply. The radial dependence would be even weaker for a Kolmogorov spectrum of the turbulence ($D(E,r)=D(E) (r/R_s)^{1/6}$). It follows that it is very difficult to use a speculative reduction in the diffusivity in the inner regions of the wind as a possible way to increase the maximum energy of the accelerated particles. 

Let us now consider the possibility that the diffusion coefficient downstream is substantially reduced with respect to the one at the shock, perhaps close to the edge of the bubble. This scenario would certainly increase the confinement time downstream, but cannot lead to a much larger value of $p_{max}$ because at some point the conditions upstream become dominant: in other words, if one could increase $p_{max}^{(2)}$ in the formalism introduced above, so that $p_{max}^{(2)}>p_{max}^{(1)}$, then the most stringent condition would be the one associated with diffusion upstream and the maximum momentum would saturate at a value closer to $p_{max}^{(1)}$. We will comment further on this point in \S \ref{ssec:spectraCR}.

Based on these considerations, we argue that the estimate of the maximum momentum derived in this section and calculated more carefully below is rather solid and that the only way to effectively increase the maximum momentum is to increase the mass loss rate and/or the wind velocity of the star cluster.

%%% ANALYTIC MODEL %%%
\section{Transport equation in the cavity and DSA at the termination shock}
\label{sec:theory}

The transport equation for non-thermal particles in the cavity can be written in spherical coordinates as follows:
\bea
  \frac{1}{r^2}\frac{\partial}{\partial r} \left[ r^2 D(r,p) \frac{\partial f}{\partial r} \right]
  - \tilde u(r) \frac{\partial f}{\partial r} + \nn \\
  + \frac{1}{r^2}\frac{d\left[ r^2 \tilde u(r) \right]}{d r} \frac{p}{3} \frac{\partial f}{\partial p} 
  -\frac{1}{p^2}\frac{\partial}{\partial p}\left[\dot p p^2 f \right]
  + Q(r,p) = 0  ,
  \label{eq:transport}
\eea
where $\tilde u(r)$ is the mean speed of the scattering centers in the shock frame, $D(r,p)$ is the diffusion coefficient, $\dot p<0$ is the rate of energy losses and the distribution function $f(r,t)$ is such that the number of particles with momentum between $p$ and $p+dp$ at the location $r$ is $4\pi p^2 f(r,p) dp$. The source term $Q(r,p)$ is an injection term, assumed here to be peaked at the injection momentum, $p_{\rm inj}$, and concentrated at the location of the termination shock. In the following we do not use it explicitly in that the TS is only introduced as a boundary condition for the solution of the transport equation and momenta are assumed to be $p\gg p_{\rm inj}$. The normalization is calculated {\it a posteriori} in terms of an efficiency of conversion of the ram pressure into pressure of non-thermal particles. This treatment is adequate insofar as non-linear feedback effects are negligible, which in turn is true provided the efficiency of particle acceleration, as a fraction of $\rho_w v_w^2$ transferred to accelerated particles, is small. 

The effective velocity $\tilde u= u + \eta  v_A$ is the sum of the plasma speed $u(r)$ and the net speed of the waves responsible for the particle scattering expressed as $\eta$ times the Alfv\'en speed $v_A$. When the parameter $\eta$ is chosen to be zero (equal number of waves moving in both directions), the effect of scattering centers disappears. Here we assume that an equal amount of waves travel in all directions upstream ($\eta=0$), while we retain the possibility to have $\eta\neq 0$ downstream. We will show below that even $\eta$ of a few percent (slight anisotropy in the propagation of waves) leads to appreciable effects on the spectrum of accelerated particles and of gamma ray emission.   

Equation~\eqref{eq:transport} is written in the assumption that stationarity is reached, applicable when the time scales for advection, diffusion and losses are appreciably shorter than the age of the star cluster that the calculation is applied to. As we discuss below, this assumption is in general satisfied but may become borderline in some cases or for some of the energies. 

Introducing the distribution function in energy $N(E,r)dE=4\pi p^2 f(r,p) dp$ and focusing on relativistic particles, $E\approx p c$, Equation~\eqref{eq:transport} can be multiplied by $4\pi p^2$, so that it becomes: 
\bea
\frac{\partial}{\partial r}\left[ \tilde u r^2 N - r^2 D \frac{\partial N}{\partial r}\right] = E\frac{\partial N}{\partial E}\left[ \frac{1}{3}\frac{d(\tilde u r^2)}{dr} - r^2 \frac{b}{E}\right] + \nn \\
+ N \left[ \frac{1}{3}\frac{d(\tilde u r^2)}{dr} - r^2 b'\right] = 0,
\label{eq:transportE}
\eea
where we omitted, for simplicity, the source term, since we are interested in $p\gg p_{\rm inj}$. Here we introduced the rate of energy losses $b(E,r)=c \dot p$, and its derivative with respect to energy $b'(E,r)=\partial b(E,r)/\partial E$.

Equation~\eqref{eq:transportE} is solved separately upstream and downstream of the TS, with the boundary condition that the particle flux at $r=0$ vanishes, namely
\beq
\left[r^2 \left(u N-D\frac{\partial N}{\partial r} \right)\right]_{r=0}=0, 
\label{eq:flux}
\eeq
and imposing a free escape boundary condition at $r=R_b$, namely $N(R_b,E)=0$. 
Moreover, it is required that the upstream and downstream solutions join continuously at $r=R_s$, namely $N_1(E,R_s)=N_2(E,R_s) \equiv N_0(E)$.

At the location of the termination shock, $r=R_s$, one can integrate Equation~\eqref{eq:transportE} across the shock discontinuity to obtain:
\beq
- D\frac{\partial N}{\partial r}|_2 + D\frac{\partial N}{\partial r}|_1 = \frac{1}{3} (\tilde u_2-\tilde u_1) E \frac{dN_0}{dE} - \frac{2}{3} (\tilde u_2-\tilde u_1) N_0,
\label{eq:TranShock}
\eeq
where $N_0(E)=N(r=R_s,E)$ is the particle distribution function at the shock and the indexes $1$ and $2$ denote, as usual, quantities evaluated immediately upstream and downstream of the TS, respectively.

Unlike in \cite{2021MNRASMorlino}, where an analytical iterative technique was used, here we solve Equation~\eqref{eq:transportE} using a mixed technique, numerical and iterative. For a given ansatz on the solution at the shock\footnote{We start from a pure power law with the expected slope, but we checked that the result does not change by adopting a different starting point.}, $N_0(E)$, Equation~\eqref{eq:transportE} is solved numerically upstream and downstream using a discretization of the grid in radius and energy discussed in Appendix \ref{app:A}. The two solutions are matched at $r=R_s$ so that $N=N_0$. Such a solution is used to determine $D\frac{\partial N}{\partial r}|_{1,2}$ as functions of energy. At this point one can introduce 
\beq
D_{1,2} \frac{\partial N}{\partial r}|_{1,2} = \Dif_{1,2}(E) \tilde u_{1,2} N_0(E),
\eeq
so that Equation~\ref{eq:TranShock} becomes:
\beq
E\frac{dN_0}{dE} = N_0 \left[ -\frac{3\Dif_2(E)}{1-\erre} + \frac{3\erre}{1-\erre}(\Dif_1(E)-1) - \frac{\erre+2}{\erre-1}\right],
\label{eq:shock}
\eeq
where we introduced the compression factor at the TS, $\erre=\tilde u_1/\tilde u_2$. The solution of Equation~\eqref{eq:shock} can be written as
\beq
N_0(E)=K E^{-\frac{\erre+2}{\erre-1}} \exp\left\{\int_0^E\frac{dE'}{E'} 
\left[-\frac{3\Dif_2}{1-\erre} + \frac{3\erre}{1-\erre}(\Dif_1-1)\right]\right\}.
\label{eq:N0}
\eeq
It is useful to notice that in the case of a plane parallel shock, $\Dif_1\to 1$ and $\Dif_2\to 0$, so that the solution reduces to the standard power law $N_0\propto E^{-\frac{\erre+2}{\erre-1}}$. The term in the exponential in Equation~\eqref{eq:N0} takes into account both the spherical symmetry and the effect of energy losses through $\Dif_1$ and $\Dif_2$. The specific energy dependence of $\Dif_{1,2}$ shapes the spectrum of accelerated particles as a result of proximity to the maximum energy and because of the spherical outflow. 

With the updated function $N_0(E)$, Equation~\eqref{eq:transportE} is solved again in the upstream and downstream regions, with the condition that for $r=R_s$ the two solutions equal the updated shape of $N_0(E)$. This iterative procedure typically requires a few iterations for convergence. 

The energy loss term in Equation~\eqref{eq:transportE} is dominated by $pp$ collisions for the energies of interest here and can be written following \cite{krakau2015} and \cite{kelner2006}:
$$
b(E,r)=5.1\times 10^{-15} K_{\pi} \, n_{\rm gas}(r) \left(\frac{E}{\rm GeV}\right)\times
$$
\beq
~~~~~~~\times\left[ 1+5.5\times 10^{-2} L+7.3\times 10^{-3} L^2\right]~\rm GeV\, s^{-1},
\eeq
where $L=\ln\left(\frac{E}{10^3\, \rm GeV}\right)$. The function $K_\pi$ has been assumed to be constant in energy, $K_{\pi}=0.13$, although a very weak energy dependence ($\sim 10\%$ on more than four orders of magnitude in energy) is found in particle codes such as Sybill and QGSJET \cite[see Fig. 2 of][]{krakau2015}. 

\section{Results}
\label{sec:results}

In this section we describe in detail the results of our calculations in terms of spectrum of particles accelerated at the TS, spectrum of CRs in the bubble and spectrum and morphology of the gamma ray emission produced in the bubble through pp collisions. Special emphasis is put on the discussion of CR energy losses, in that this phenomenon affects both the spectrum of escaping CRs and the interpretation of gamma ray emission, and was not accounted for in previous works on the topic.    

\subsection{Spectrum of accelerated particles}
\label{ssec:spectraCR}

Our benchmark case in terms of choice of parameters is that of the Cygnus OB2 Cocoon, for which we assume $v_w=2800$ km/s, $\dot M=1.5\times 10^{-4}M_\odot~ \rm yr^{-1}$, age of $3$ million years, and density of the outside ISM $20~\rm cm^{-3}$. With these values of the parameters, the luminosity of the star cluster is $L_w=3.8\times 10^{38}$ erg/s. The termination shock is located at $R_s=15.3~\rm pc$ while the outer edge of the bubble is at $96$ pc. While this set of parameters defines our benchmark model, in the following we will investigate the effect of changing these numbers within a reasonable range that may describe a more generic star cluster or account for uncertainties in the value of these parameters for the case of Cygnus OB2 \citep{Menchiari2023}. 

\begin{figure}
\centering
\includegraphics[width=.45\textwidth]{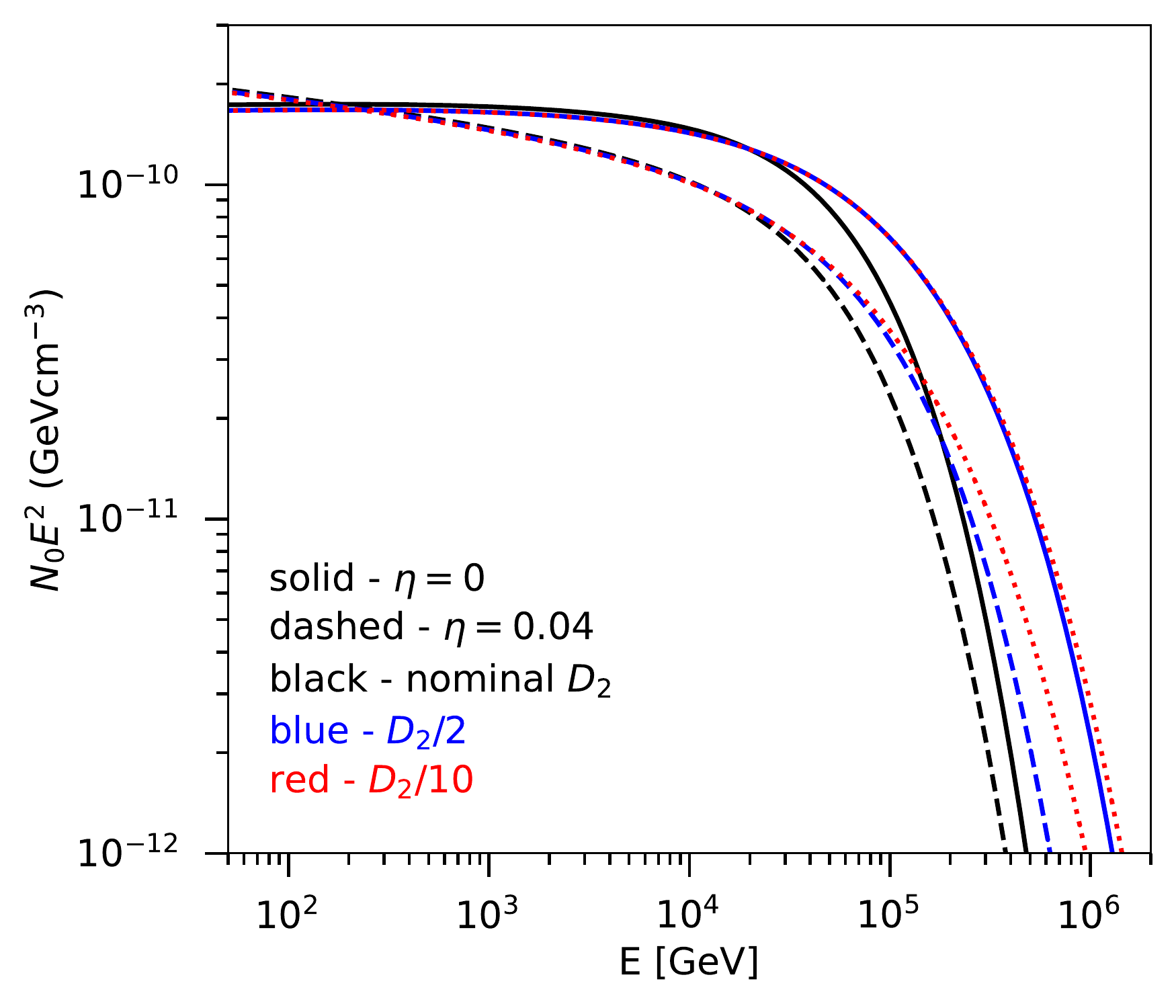}
\caption{Spectrum of accelerated particles at the TS for $\tilde u_2=u_2+\eta v_{A,2}$ with $\eta=0$ (solid lines) and $\eta=0.04$ (dashed lines). An efficiency of CR acceleration $\xi_{CR}=0.01$ has been used. The black curves refer to the nominal diffusion coefficient downstream, while the blue and red curves have been obtained by suppressing artificially $D_2$ by a factor $2$ and $10$ respectively. The low energy slope of $N_0(E)$ for $\eta=0.04$ (dashed lines) is $2.08$.}
\label{fig:CRspec}
\end{figure}

In Figure~\ref{fig:CRspec} we show the spectrum of accelerated particles at the TS in our benchmark case, assuming that the scattering waves downstream of the TS are fully isotropic ($\tilde u_2=u_2$, solid black line) or that alternatively there is a 4\% excess of waves moving away from the shock toward downstream ($\tilde u_2=u_2+0.04 v_{A,2}$, dashed black line). The latter case is expected to lead to a steeper spectrum \cite[]{Bell:1978p1342}  \cite[see also discussion in][]{2021MNRASMorlino}. These curves are obtained using the Kraichnan expressions for the diffusion coefficients described in \S~\ref{sec:bubble}, with $\eta_B=0.1$. The blue (red) curves for each of the two cases listed above show the spectra of accelerated particles in the case that the diffusion coefficient downstream is artificially reduced by a factor 2 (10) to mimic the excitation of MHD instabilities behind the shock front \cite[]{Giacalone:2007p962}. 

Figure~\ref{fig:CRspec} illustrates in a clear way how tricky is the definition of the maximum energy in the spherical geometry typical of a stellar cluster: although for the parameters that we have chosen here the maximum energy can be easily read off Figure~\ref{fig:Pmax} to be of order $\sim 1$ PeV, one can see that the spectrum of particles accelerated at the TS starts dropping appreciably at energy $\lesssim 100$ TeV, while in the PeV region the spectrum is already rapidly dropping. As discussed by \cite{2021MNRASMorlino}, this effect is due to the appearance of a sort of mean plasma speed upstream: for low energies, this effective speed is close to $v_w$ and the spectrum is the same that one would obtain for a plain shock. At high energies, when the diffusion length upstream is not negligible compared with the radius of the TS, the effective speed becomes less than $v_w$, which implies a smaller effective compression factor and a steeper spectrum. This effect is more pronounced for weak energy dependence of the diffusion coefficient: for Kolmogorov scaling, one has a gradual steepening rather than a cutoff, that starts already in the TeV region. For Bohm diffusion the spectrum would start cutting off at approximately the maximum energy. However, as discussed above, this case seems to be poorly justified at least for the parameters of Cygnus OB2. The case of Kraichnan scaling adopted here is somewhat intermediate between the Bohm and the Kolmogorov cases. 

The blue and red curves have been introduced to comment on the effect that decreasing the diffusion coefficient downstream with respect to the nominal value would have on the maximum momentum of accelerated particles. Reducing $D_2$ by a factor 2 does indeed lead to a slight increase in $p_{\max}$, as expected; however an additional reduction, to $D_2/10$ does not lead to any appreciable change, except perhaps a small deviation in the shape of the cutoff. This is because for a too small value of the downstream diffusion coefficient, it is the upstream confinement that becomes more constraining in terms of maximum momentum. This can be easily appreciated by comparing the blue and red curves in Fig. ~\ref{fig:CRspec}.

\begin{figure}
\centering
\includegraphics[width=.45\textwidth]{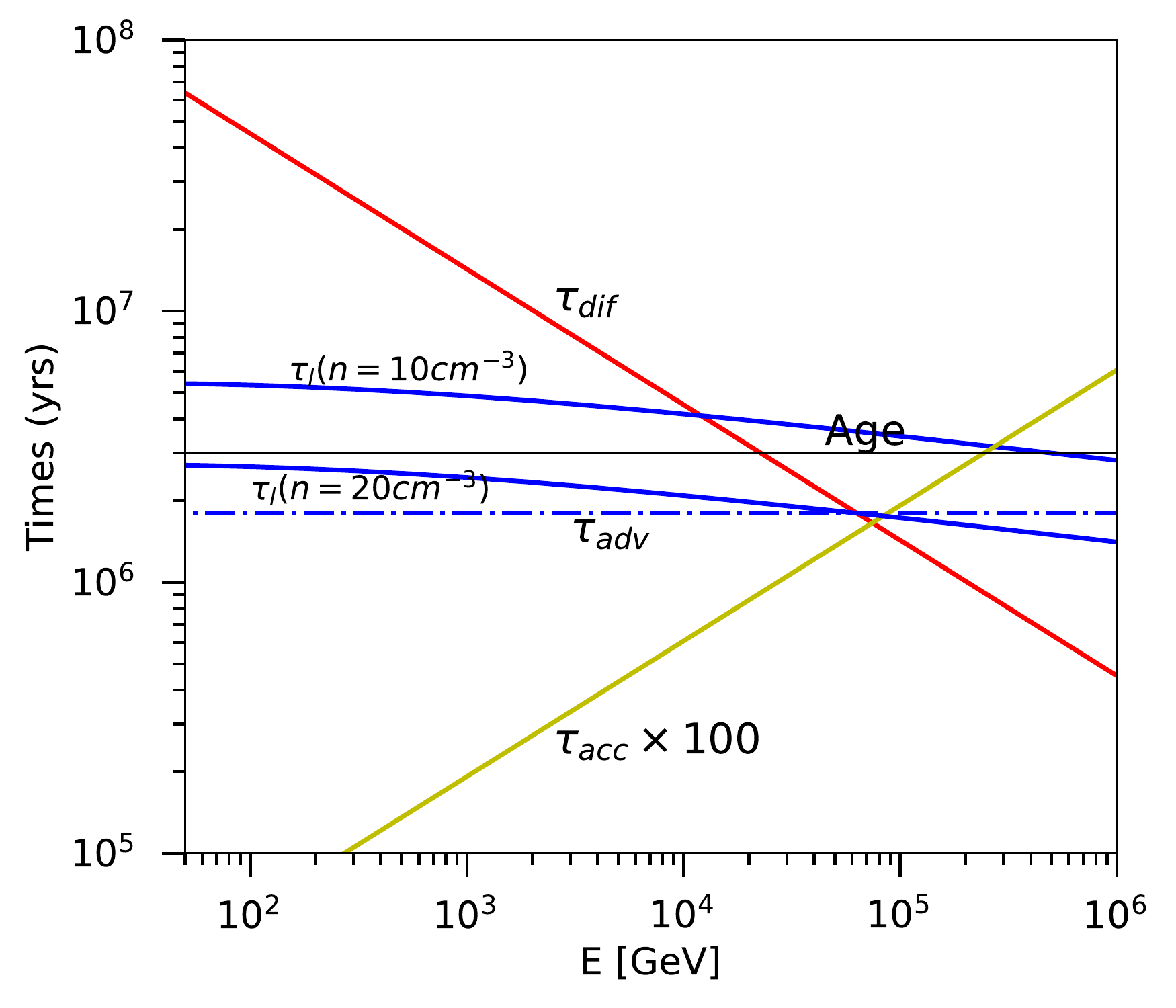}
\caption{Time scales for diffusive escape from the bubble ($\tau_{\rm dif}$), advection ($\tau_{\rm adv}$), losses ($\tau_l$) for two values of the gas density in the cavity ($n=10$ and $n=20~\rm cm^{-3}$), and acceleration ($\tau_{\rm acc}$), multiplied here by 100 to make it visible in the same plot. These timescales are compared with the age of the star cluster (solid black line).}
\label{fig:Times}
\end{figure}

So far we have not discussed the role of energy losses: this is because the time scale for losses is much longer than the acceleration time, hence the spectrum at the shock is weakly affected by losses. However this does not mean that losses are unimportant, in that they affect the spatial distribution and the spectrum of particles in the downstream region. This is illustrated in Figure~\ref{fig:Times}, where we compare the diffusion time scale in the region downstream of the shock, $\tau_{\rm dif}(E)=R_b^2/D_2(E)$, the time scale of advection
\beq
\tau_{\rm adv} = \int_{R_s}^{R_b} \frac{dr}{u(r)}=\frac{1}{3}\frac{R_b}{u_2}\left( \frac{R_b}{R_s}\right)^2\left[1-\left( \frac{R_s}{R_b}\right)^3
\right] \;,
\eeq
the timescale for losses due to pion production, $\tau_l=E/b(E)$ (for target density $n=10\, \rm cm^{-3}$ and $n=20\, \rm cm^{-3}$)  and the age of the star cluster, $\tau_{\rm age}$. The acceleration time, approximated here as 
\beq
\tau_{\rm acc}(E)\approx \frac{3}{\tilde u_1-\tilde u_2} \left[\frac{D_1}{\tilde u_1}+\frac{D_2}{\tilde u_2} \right],
\eeq
is shown multiplied by a factor 100 in order to make it visible in the same plot. Clearly the maximum energy is not limited by energy losses in a star cluster resembling Cygnus OB2. 

While losses are not fast enough to shape the spectrum of accelerated particles at the shock location, they can modify the particle spectrum in the downstream region and, as a consequence, the spectrum of particles escaping the cavity as well. Notice that the observed gamma ray emission mainly comes from the region downstream of the TS (see \S~\ref{ssec:gamma}), hence it carries information on the effect of energy losses as well. 

\begin{figure}
\centering
\includegraphics[width=.45\textwidth]{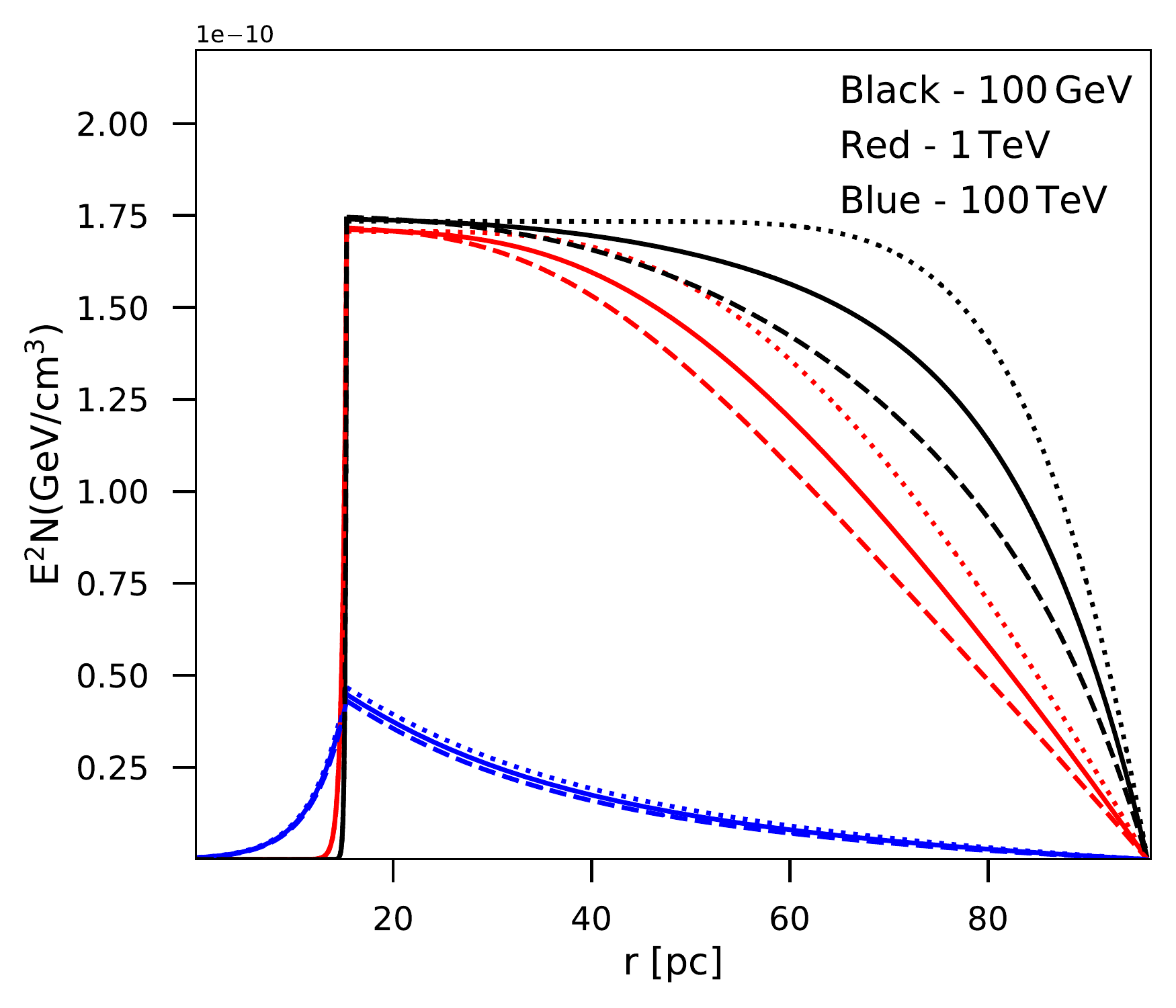}
\caption{Spatial distribution of accelerated particles upstream and downstream of the TS, for three energies (as indicated) and for density $n=10^{-3}\, \rm cm^{-3}$ (dotted lines), $n=10\, \rm cm^{-3}$ (solid lines) and $n=20\, \rm cm^{-3}$ (dashed lines). All curves refer to the simple case of absence of Alfv\'enic drift downstream, namely $\eta=0$.}
\label{fig:Space}
\end{figure}

In Figure~\ref{fig:Space} we show the spatial distribution of particles with energy $100$ GeV, $1$ TeV  and $100$ TeV for a low target density in the bubble, $n=10^{-3}\, \rm cm^{-3}$ (where no appreciable losses are expected) and for the more realistic values of $n=10\, \rm cm^{-3}$  and $n=20\, \rm cm^{-3}$ (the gas is expected to be mainly in the form of clumped neutral Hydrogen, as discussed in \S~\ref{sec:clumps}). One can clearly identify the position of the TS at $\sim 15$ pc from the center and the edge of the bubble at $\sim 96$ pc. From Figure~\ref{fig:Times} one can appreciate that the transport in the downstream region is mainly regulated by advection and losses for $E\lesssim 50$ TeV, while diffusion plays the most important role at higher energies. Hence the effect of losses is most visible in the spatial distribution of lower energy particles in the downstream region (black and red lines in Figure~\ref{fig:Space}). At 100 TeV, losses do not lead to an appreciable change in the spatial distribution of the accelerated particles, although losses may become important at such energies if larger values of the density in the bubble are assumed. 

In the upstream region, as expected, there is no appreciable difference between the three situations, due to the fact that losses are too slow to operate in one advection time over a diffusion length of particles of given momentum. Downstream of the TS the difference between the case with weak or no losses, as previously investigated by \cite{2021MNRASMorlino}, and the ones with losses (solid and dashed lines) is rather remarkable. Clearly this difference reflects on the volume integrated gamma ray emission, especially in terms of the efficiency required for particles acceleration: the fact that pp energy losses are only weakly dependent on energy does not lead to specific spectral breaks, but rather to a global change of normalization, although small effects are present due to the relative importance of other effects (diffusion versus advection and losses). 

The difference in normalization at different energies in Figure~\ref{fig:Space} reflects the decreasing behavior of the spectrum of accelerated particles as a function of energy, and eventually the appearance of the suppression due to approaching the maximum energy. Here we considered the case in which $\tilde u_2=u_2$, namely $\eta=0$. If $\eta>0$, the main change appears in the different normalization due to the steeper spectrum of accelerated particles. 

\subsection{Gamma ray emission from Cygnus OB2: spectra and morphology}
\label{ssec:gamma}

In this section we discuss how the calculations of the spectrum and spatial distribution of accelerated particles affect the gamma ray emission observed from the direction of the Cygnus cocoon, where observations have recently become available both in the GeV \citep{ackermann-Cygnus2011} and the TeV energy range \cite[]{2021NatureHAWC}. 

Before trying to provide a physical description of the gamma ray emission from the Cygnus OB2 association, it is important to discuss in some detail what the observations actually refer to: the HAWC telescope observed the central $\sim 2$ degrees of the region, corresponding to about $\sim 50$ pc at a distance of 1.4 kpc appropriate for the Cygnus region. This region was divided in four annuli and for each of them the gamma ray emission was measured. The total flux from the Cygnus region, including the region outside the central $\sim 2$ degrees, was estimated assuming a gaussian spatial distribution in two dimensions (distribution in the sky of an extended source). This simple assumption implies that the flux of gamma rays that is actually observed (within the central region) accounts for about 39\% of the total emission, while 61\% of the total emission is contributed by the outside region, which is not directly observed. Clearly this assumption would be inconsequential if the spatial distribution of the gamma ray emission were really Gaussian and if the spectrum of the gamma ray emission were the same throughout the emitting region. Unfortunately, as we discuss below and as we illustrated above in terms of CR spatial distribution, both these assumptions are violated, which makes it difficult to extract information on the details of the acceleration and transport of particles in the cocoon. There are however several pieces of information that can be inferred from available data: first, we point out that the gamma ray spectrum is available for each of the four annuli discussed above (Binita Hona, private communication), hence one could restrict the analysis to the comparison between our predictions and the data in the same annuli. This comparison is however possible only in the $\sim 1-100$ TeV range, while at present the same information is not readily available in the GeV range (Fermi-LAT). On the other hand, the total gamma ray flux (with a similar assumption of Gaussian spatial distribution) has been published by the Fermi-LAT collaboration \citep{ackermann-Cygnus2011} while the flux per annuli (but not the spectrum) has been obtained by \cite{Aharonian+2019NatAs}. 
We also notice that \cite{Aharonian+2019NatAs} provide an estimate of the total gas content inside the Cygnus region per annuli which translates to an average target density of $40-100\, \rm cm^{-3}$. Such an estimate might be overestimated if the detected gas along the line of sight is not all concentrated in the Cygnus region, hence, in the calculations below we adopt a more conservative estimate of $10-20\, \rm \rm cm^{-3}$.

The second avenue to gather information on the physics of this system is to limit our calculation to the same central region and artificially adopt the same Gaussian assumption on the spatial distribution in order to estimate the total flux from the whole region, in a way that our flux is as consistent as possible with the procedure adopted in observing the Cygnus region. This would allow us to quantify the implications of the assumption of Gaussian spatial distribution of the gamma ray emission and its impact on the evaluation of the parameters of the system. 

\begin{figure*}
\centering
\includegraphics[width=.45\textwidth]{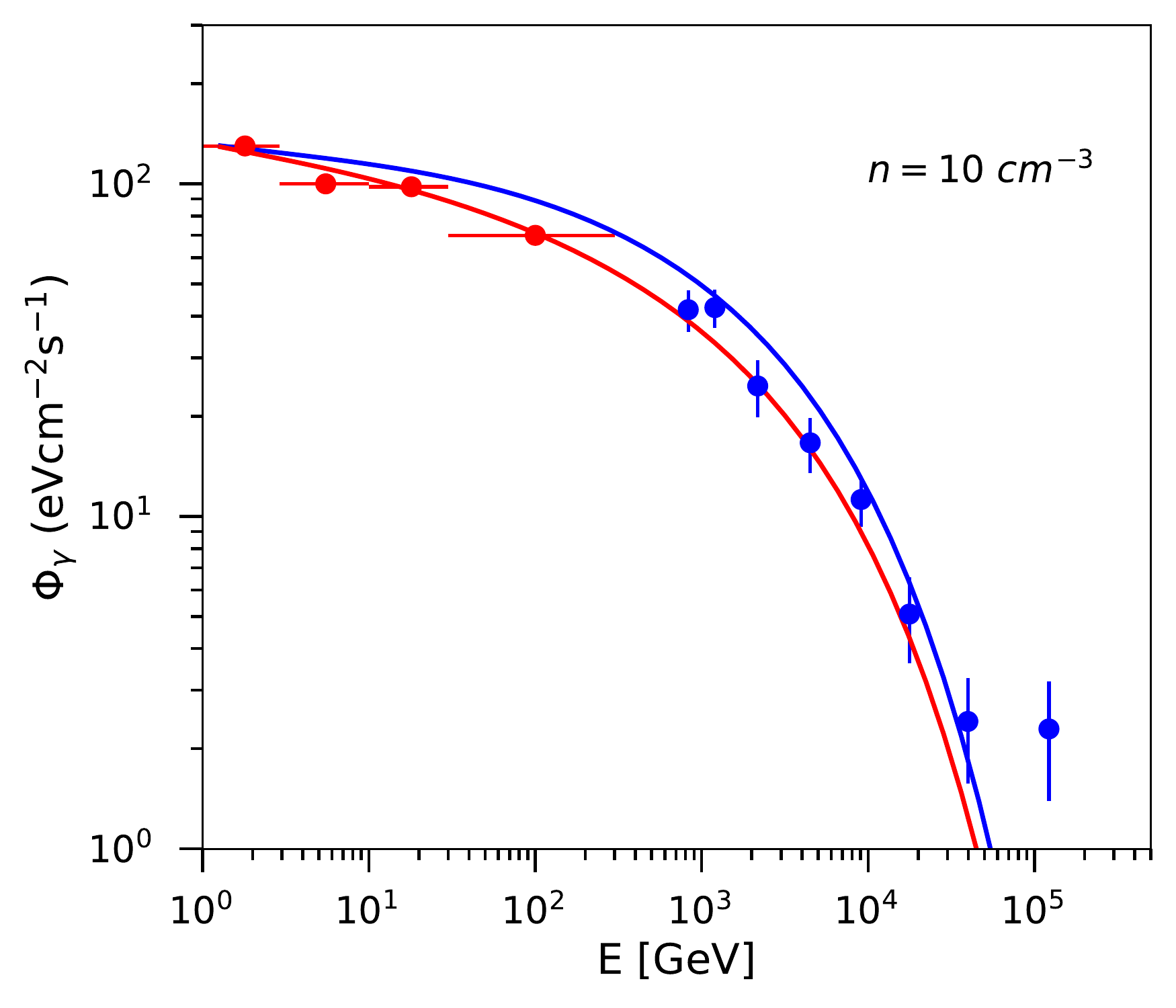}
\includegraphics[width=.45\textwidth]{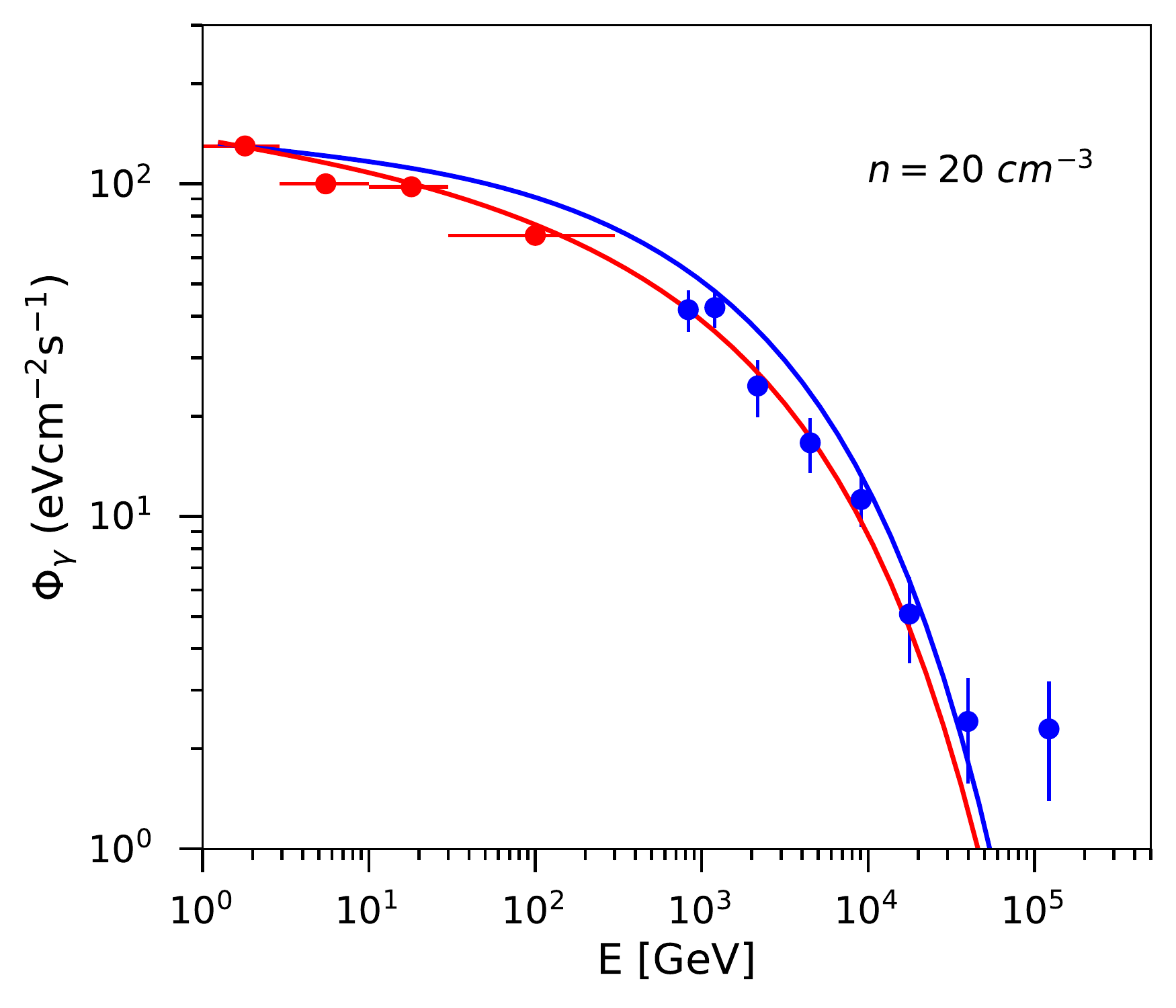}
\caption{Comparison between the gamma-ray flux emitted from the entire wind-blown bubble (red lines) with the emission from the innermost 55 pc rescaled assuming a Gaussian distribution of the emission (blue lines) for the case of Cygnus Cocoon. Left and right panels differ for the assumed gas density, being $n=10~\rm cm^{-3}$ and $n=20~\rm cm^{-3}$, respectively. The CR acceleration efficiency has been chosen so as to fit the flux in the lowest energy bin for the blue line: $\xi_{\rm CR}=1\%$ in the left panel and $0.55\%$ in the right panel. The red lines have efficiency $\xi_{\rm CR}=1.3\%$ in the left panel and $\xi_{\rm CR}=0.77\%$ in the right panel. Data are from Fermi-LAT and HAWC.}
\label{fig:TwoWays}
\end{figure*}

Before trying to obtain a viable description of the gamma ray emission from the Cygnus OB2 region, it is instructive to show the effect of integrating on the inner region and assuming Gaussian distribution versus integrating over the whole volume of the bubble.  
The gamma ray emission from a given line of sight at distance $y$ from the center of the star cluster is calculated as 
\beq
  I_\gamma(E_\gamma,y) = 2 \int_0^{l_{\max}}  J_\gamma(E_\gamma,l) \, dl\,,
\label{eq:line}
\eeq
where $J_\gamma$ is the gamma ray emissivity per unit volume, per unit time and unit energy at energy $E_\gamma$, at the location along the line of sight $l$ related to the radial distance from the center through $r^2-y^2=l^2$. Here $l_{\max}=\sqrt{R_b^2-y^2}$. The emissivity is calculated through the standard formalism using the cross sections by \cite{kelner2006}. The factor 2 in Eq. \eqref{eq:line} is due to the symmetry of the system. The total gamma ray flux can then be calculated as
\beq
\Phi_\gamma(E_\gamma) = E_\gamma^2 \int_0^{R_b} I_\gamma(E_\gamma,y) 2 \pi y dy.
\label{eq:FluxGamma}
\eeq

In Figure~\ref{fig:TwoWays} we illustrate the effect of the two different ways of calculating the gamma ray flux from the direction of Cygnus: both curves are obtained for $\eta=0$ (no net Alfv\'en speed) and for $D_2$ due to pure compression. The blue curves show the gamma ray flux obtained by integrating the emission within the central 55 pc \cite[as done in][]{2021NatureHAWC} and then renormalizing the flux to account for a 2D Gaussian spatial distribution. In the left (right) panel the gas density has been chosen as $n=10 \, \rm cm^{-3}$ ($n=20\,\rm cm^{-3}$). In the lower density case the efficiency of acceleration is required to be $\xi_{\rm CR}=1\%$, while it is $\xi_{\rm CR}=0.55\%$ for $n=20 \, \rm cm^{-3}$. The inferred efficiency does nor scale exactly as the inverse of the density because of energy losses.

Clearly the effect becomes much more pronounced for star clusters with larger mean densities. The red curves in the two plots of Figure~\ref{fig:TwoWays} represent the gamma ray fluxes obtained by integrating over the whole volume of the bubble. The most immediate implication of the comparison between the red and blue curves is that the spectral shape in the low energy region is different. Moreover, the larger CR acceleration efficiencies required with respect to the blue lines ($\xi_{CR}=1.3\%$ on the left panel and $\xi_{\rm CR}=0.77\%$ in the right panel) shows that the assumption of Gaussian spatial distribution leads to overestimating the flux of gamma rays, in addition to providing different information of the spectrum of the accelerated particles. In the following, the gamma ray fluxes will be assumed to be calculated by integrating the emission from the central region and renormalizing to account for the 2D Gaussian, a procedure that seems the closest to the one adopted by the HAWC Collaboration in showing their results. 

\begin{table*}
\begin{center}
\begin{tabular}[c]{l c c c c c c c c}  
\toprule \toprule
Models  &  $\dot{M}$ & $v_w$ & $n_0$ & $\eta_B$ & $n_2$ & $\eta$ & $D_1/D_2$ \\
  & [$M_{\odot}\, \rm yr^{-1}$] & [km/s] & [cm$^{-3}$] &  & [cm$^{-3}$] &  &  \\
\midrule
Model 1 & $1.5 \times 10^{-4}$ & 2800 & 20 & 0.1 & 10\;(20) & 0   & $11^{1/4}$  \\
Model 2 & $1.5 \times 10^{-4}$ & 2800 & 20 & 0.1 & 10\;(20) &  0.04 & $11^{1/4}$  \\
Model 3 & $1.5 \times 10^{-4}$ & 2800 & 20 & 0.1 & 10\;(20) & 0.04 & $2 \times 11^{1/4}$  \\
Model 4 & $2.0 \times 10^{-4}$ & 3000 & 20 & 0.1 &  10\;(20) &  0.04   & $11^{1/4}$  \\
\bottomrule %\bottomrule
\end{tabular}
\caption{Parameters' values for the models discussed in the text. The ratio $D_1/D_2 = 11^{1/4}$ applies to the case of pure compression and results from the assumptions of randomly oriented magnetic field, so that $B_2 = \sqrt{11}\, B_1$, and of Kraichnan diffusion, where $D\propto B^{-1/2}$. The other parameter are the same for all models: $t_{\rm age} = 3 \,\rm Myr$ and $n_{\rm ISM}= 20 \, \rm cm^{-3}$}.
\label{tab:Models}
\end{center}
\end{table*}

\begin{figure*}
\centering
\includegraphics[width=.45\textwidth]{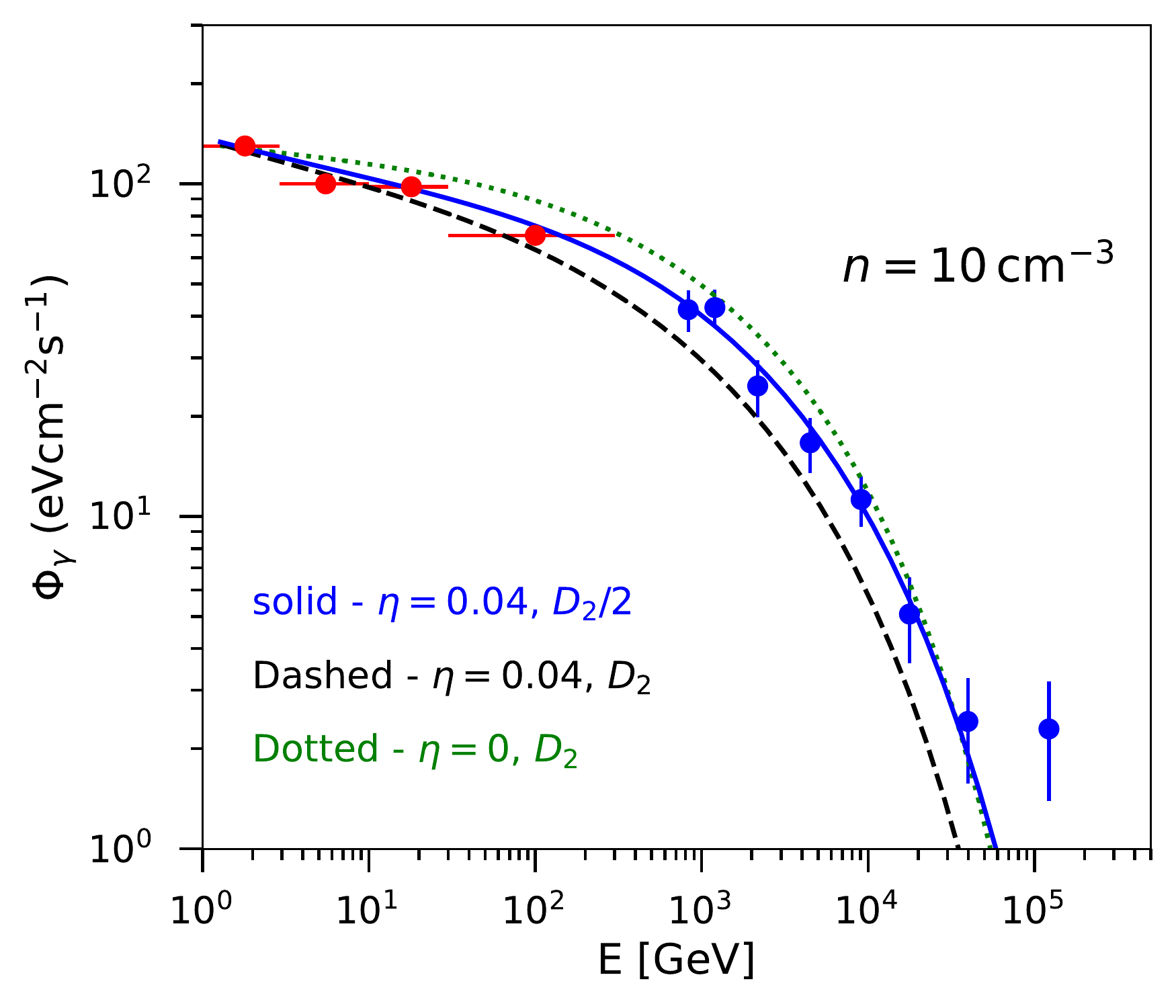}
\includegraphics[width=.45\textwidth]{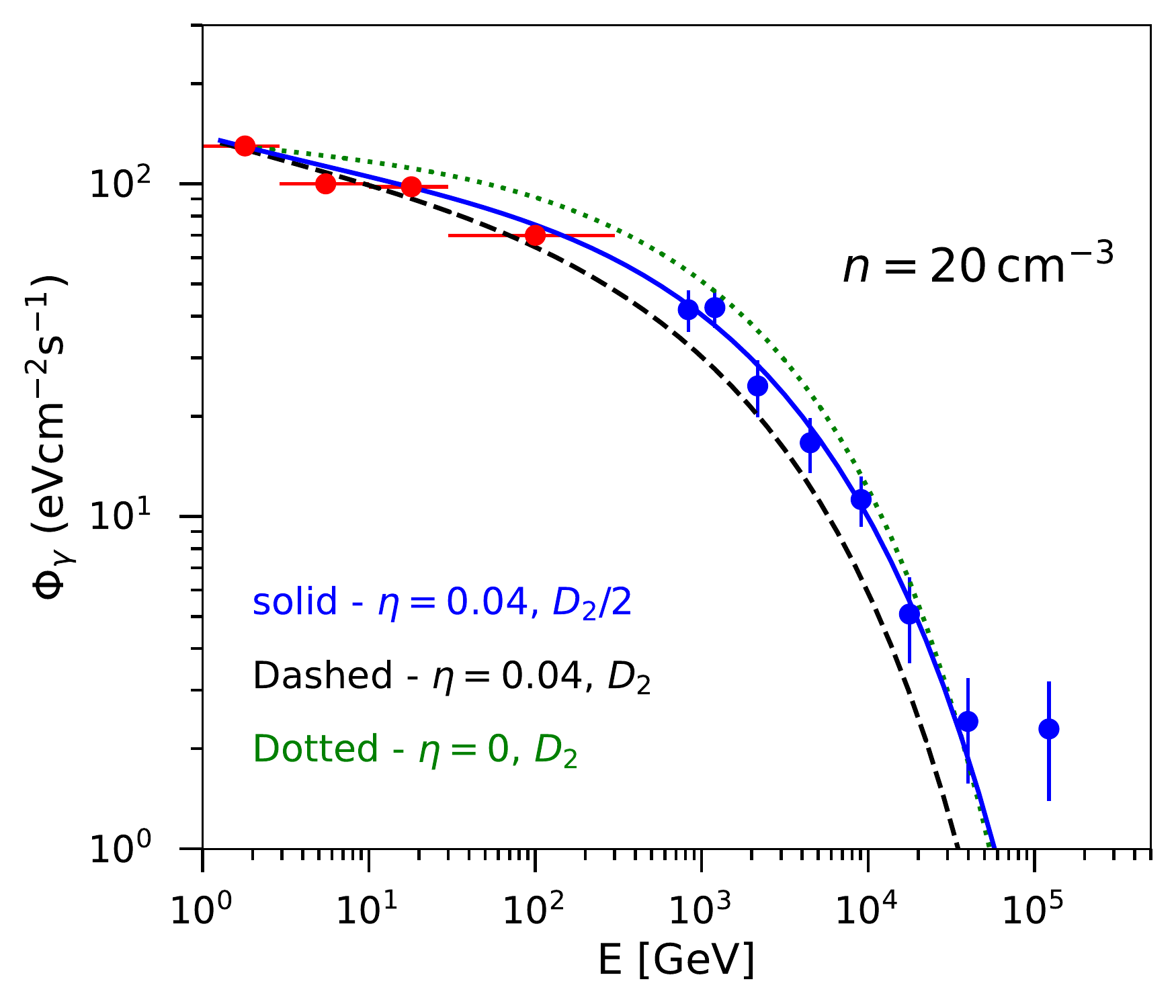}
\caption{Volume integrated gamma ray flux for Models 1, 2 and 3 described in the text and reported in Table~\ref{tab:Models}. The gas density in the bubble is $n=10\,\rm cm^{-3}$ in the left panel and $n=20\,\rm cm^{-3}$ in the right panel. The CR acceleration efficiency has been changed to obtain a best fit to the flux in the lowest energy bin, and varies between $0.45\%$ and  $\sim 1\%$.}
\label{fig:GammaGen}
\end{figure*}

In Figure~\ref{fig:GammaGen} we show the gamma ray emission from the Cygnus region for the benchmark values of the parameters of the wind and for three different models of CR transport:  Model 1) $\eta=0$ and $D_2$ due only to shock compression; Model 2) $\eta=0.04$ and $D_2$ as in Model 1; Model 3) $\eta=0.04$ and $D_2$ suppressed artificially by a factor 2. For these three models we discuss the cases of the density in the bubble $n=10 \, \rm cm^{-3}$ and $n=20\, \rm  cm^{-3}$, which imply different levels of energy losses and, as a consequence, different CR acceleration efficiency (see Table~\ref{tab:Models} for a summary of the parameters).

Model 1 (dotted green line) leads to exceedingly hard spectra in the low energy region that fail to reproduce the data. Model 2 (black lines) clearly leads to slightly steeper CR spectra (the low energy slope is $2.08$) and provides a better description of the spectrum of the gamma ray emission in the Fermi-LAT energy region. 
On the other hand, the gamma ray emission cuts off at too low energies and fails to reproduce the data for $E_\gamma\gtrsim 1$ TeV. Notice that for this specific case the maximum energy estimated as discussed in Sec. \ref{sec:Pmax} is $\gtrsim$ PeV. Nevertheless, for Kraichnan spectrum of the turbulence the shape of the CR spectrum manifests a gradual steepening already at $E\ll E_{\max}$, causing a corresponding steepening of the gamma ray spectrum. This problem becomes even more severe for Kolmogorov turbulence. 

Finally, Model 3 (solid blue lines) better agrees with data also at higher energies, due to the fact that in this model the downstream diffusion coefficient was reduced by a factor 2 to mimic the possible magnetic field amplification due to hydro-dynamical instabilities, thereby causing a somewhat higher maximum energy. For both cases of $n=10 \, \rm cm^{-3}$ and $n=20 \, \rm cm^{-3}$ this situation leads to a satisfactory description of the data, requiring a CR acceleration efficiency below $1\%$. This low efficiency justifies the decision of neglecting the non-linear dynamical feedback in particle acceleration at the termination shock. 

\begin{figure}
\centering
\includegraphics[width=.45\textwidth]{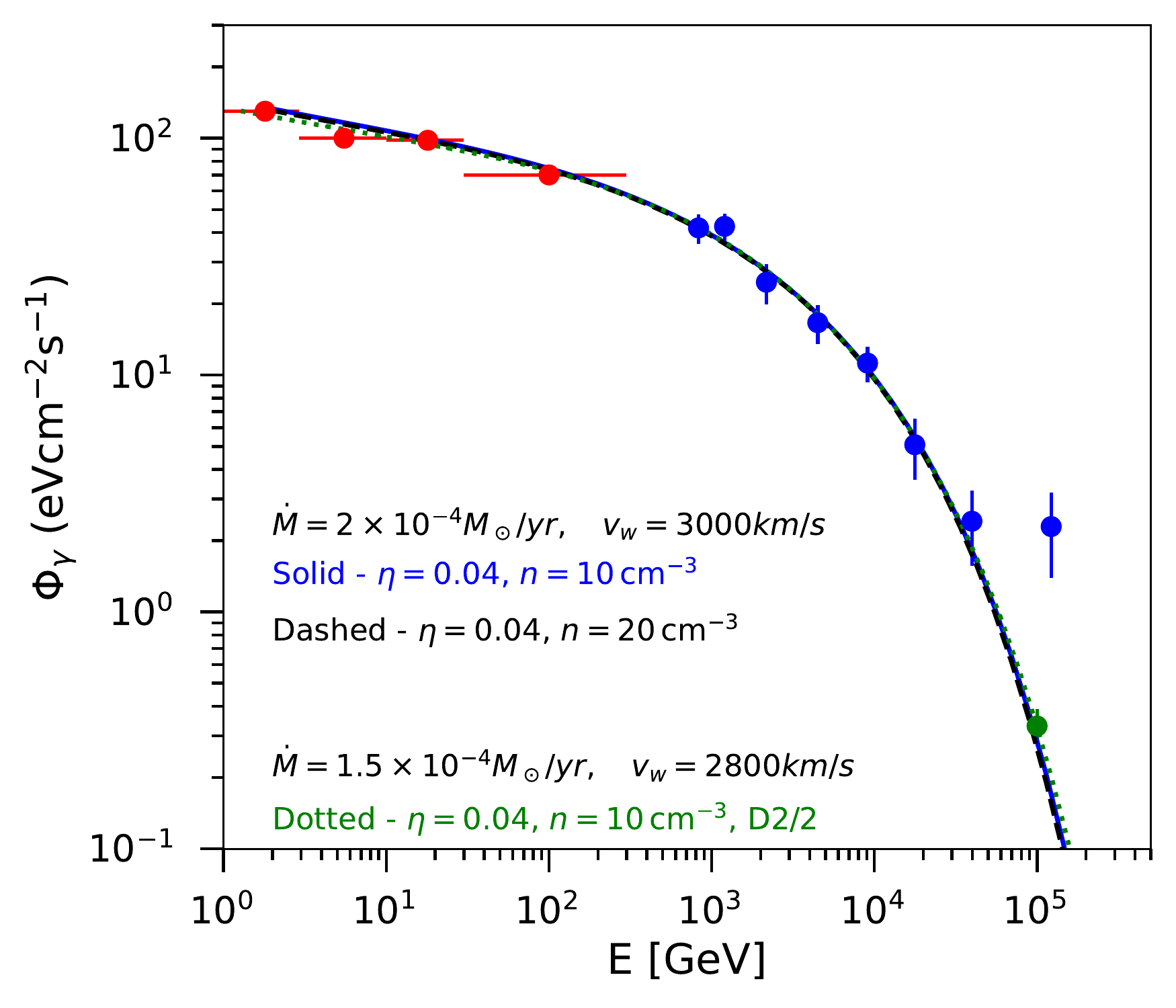}
\caption{Volume integrated gamma ray flux for Model 4, with $\eta=0.04$ and $n=10$ (solid) and $20~\rm cm^{-3}$ (dashed), and for Model 3 with $n=10~\rm cm^{-3}$ (dotted). The CR acceleration efficiency in the three cases is $\xi_{\rm CR}=0.69\%$, $\xi_{\rm CR}=0.37\%$ and $\xi_{\rm CR}=0.81\%$ respectively. The preliminary LHAASO data point \citep{2021Natur.594...33C} has also been introduced.}
\label{fig:GammaFit}
\end{figure}

It is worth stressing that the total gamma-ray emission alone cannot constrain all the parameters of the model. In fact there is a degeneracy between wind luminosity, acceleration efficiency and gas density. For instance, in Figure~\ref{fig:GammaFit} we show three possible sets of parameters for the wind and for CR transport that allow equally good and basically indistinguishable descriptions of the spatially integrated gamma ray flux: two lines refer to the more energetic Model 4 (see Table~\ref{tab:Models}) with target density $n=10$ (solid blue line) and $n=20\,\rm cm^{-3}$ (dashed black line); the third case refers to Model 3, where the diffusion coefficient downstream is reduced artificially by a factor 2 (dotted line). All these cases lead to somewhat larger maximum energy so that the gamma ray spectrum extends to higher energies and, in addition to providing a good description of HAWC data, also explain the preliminary integral flux above $100$ TeV recently quoted by LHAASO \citep{2021Natur.594...33C} (green data point at 100 TeV in Figure~\ref{fig:GammaFit}).

A note of caution is in order: the comparison between the spectra of gamma ray emission claimed by different experiments is a delicate issue, in that slightly different size of the region of interest and the assumption of the spatial profile outside this region may lead to deformation in the spectrum and in turn to inferring incorrect values of physical parameters. This argument applies especially well to the LHAASO data point, for which only preliminary information is available at present. It is of crucial importance to await for more details of the gamma ray spectrum and morphology at $E_\gamma\gtrsim 100$ TeV to build a more physical picture of CR transport in the Cygnus region. 

Based on the information available at the present time, the gamma ray emission at $\sim 100$ TeV clearly indicates the presence of some accelerated particles in the PeV range. However, this does not imply that the Cygnus OB-2 is necessarily a PeVatron from the point of view of explaining the CR spectrum, in that one can clearly see that the effective maximum energy (defined as the energy where the power law extrapolation from lower energies in the CR spectrum drops by $1/e$) is well below PeV (see also Figure~\ref{fig:CRspec}). The experimental facilities that are being built are reaching such a high sensitivity that they are now able to measure the flux in the cutoff region down to very low fluxes. This result is in fact of the utmost importance, in that the shape of the cutoff carries information about the acceleration process. However, detection of ~100 TeV photons does not automatically imply that a source is able to produce enough protons at $\sim$PeV energy so as to explain the {\it knee}. 

\begin{figure}[t]
\centering
\includegraphics[width=.45\textwidth]{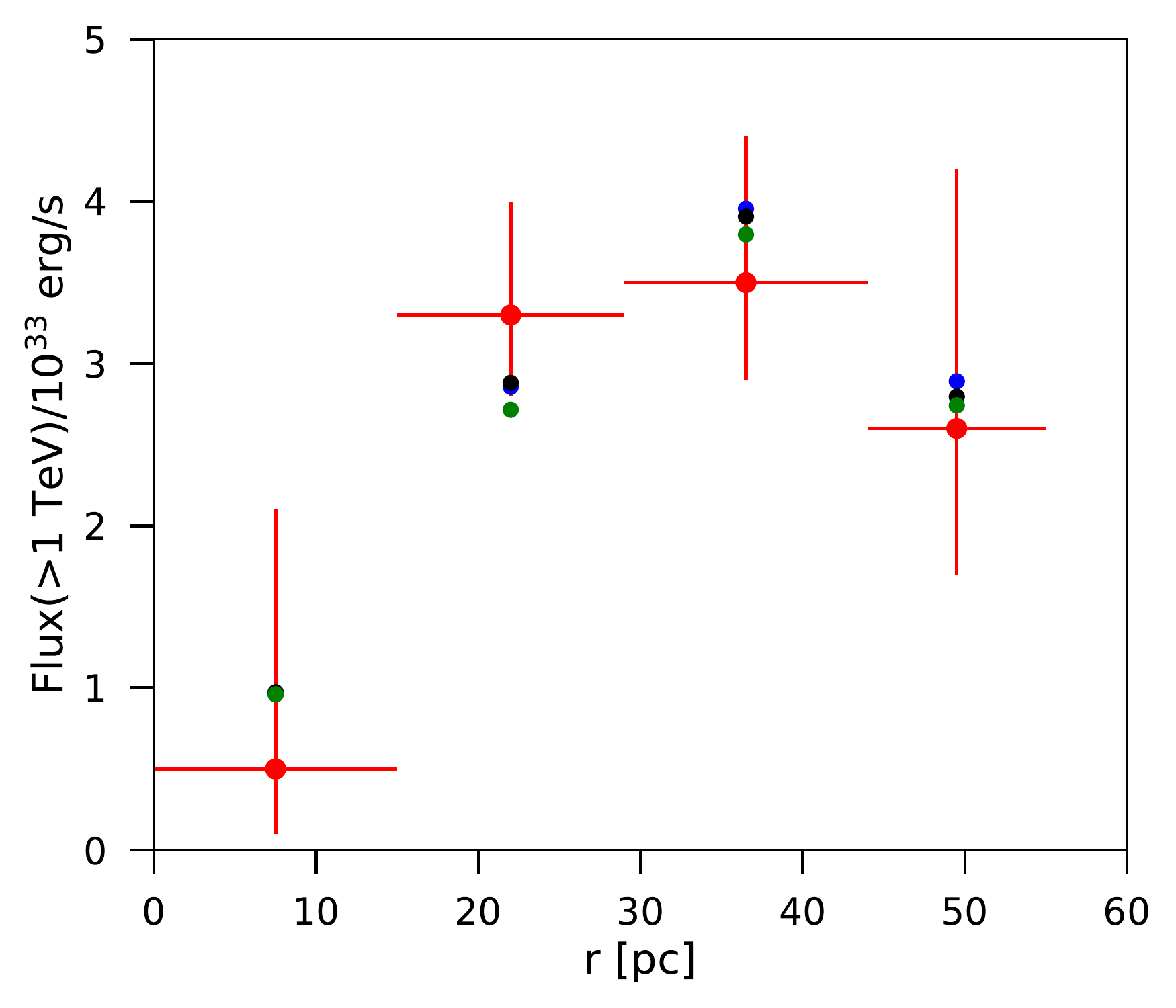}
\caption{Gamma ray flux above $1$ TeV in rings around the center of Cygnus OB2. The red dots with error bars are the results obtained by HAWC \citep{2021NatureHAWC}. The colored circles refer to the same cases of Figure~\ref{fig:GammaFit}: Model 4 with $n=10~\rm cm^{-3}$ (blue dots) and $n=20~\rm cm^{-3}$ (black dots); Model 3 with $n=10~\rm cm^{-3}$ and the downstream diffusion coefficient reduced by a factor 2 (green dots). In all cases $\eta=0.04$.}
\label{fig:GammaMorph}
\end{figure}

An additional piece of information on the origin of the accelerated particles and of the non-thermal emission is carried by the morphology of the gamma ray emission. The flux of gamma rays observed by HAWC with energy $>1$ TeV is shown in Figure~\ref{fig:GammaMorph} in four bins with increasing distance from the center of the star cluster. These fluxes are obtained by integrating the gamma ray flux in rings around the center of the cocoon \citep{2021NatureHAWC}. The results of our calculations for the same cases illustrated in Figure~\ref{fig:GammaFit} (and reported in Table~\ref{tab:Models}) are shown as thick dots.

\begin{figure*}
\centering
\includegraphics[width=.45\textwidth]{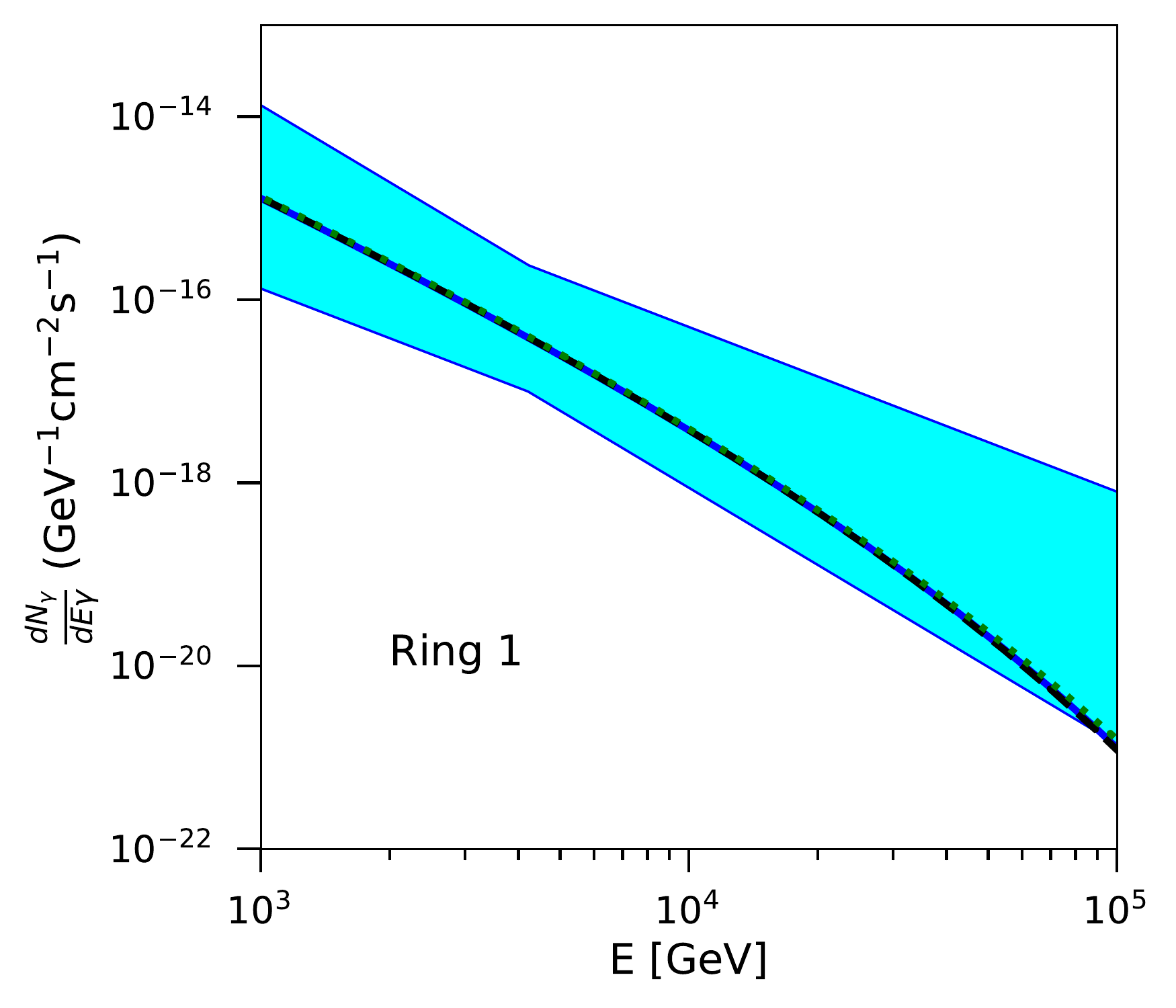}
\includegraphics[width=.45\textwidth]{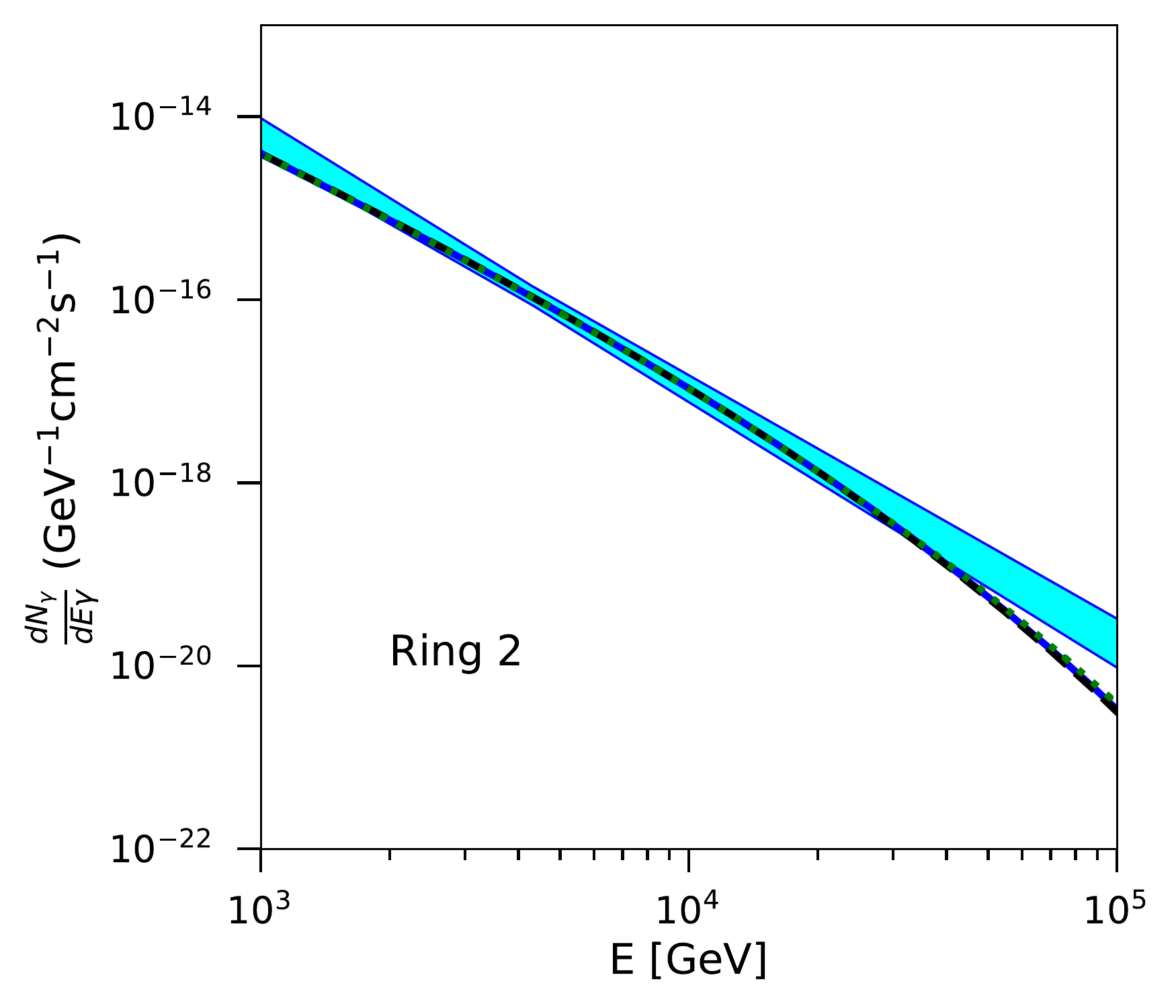}
\includegraphics[width=.45\textwidth]{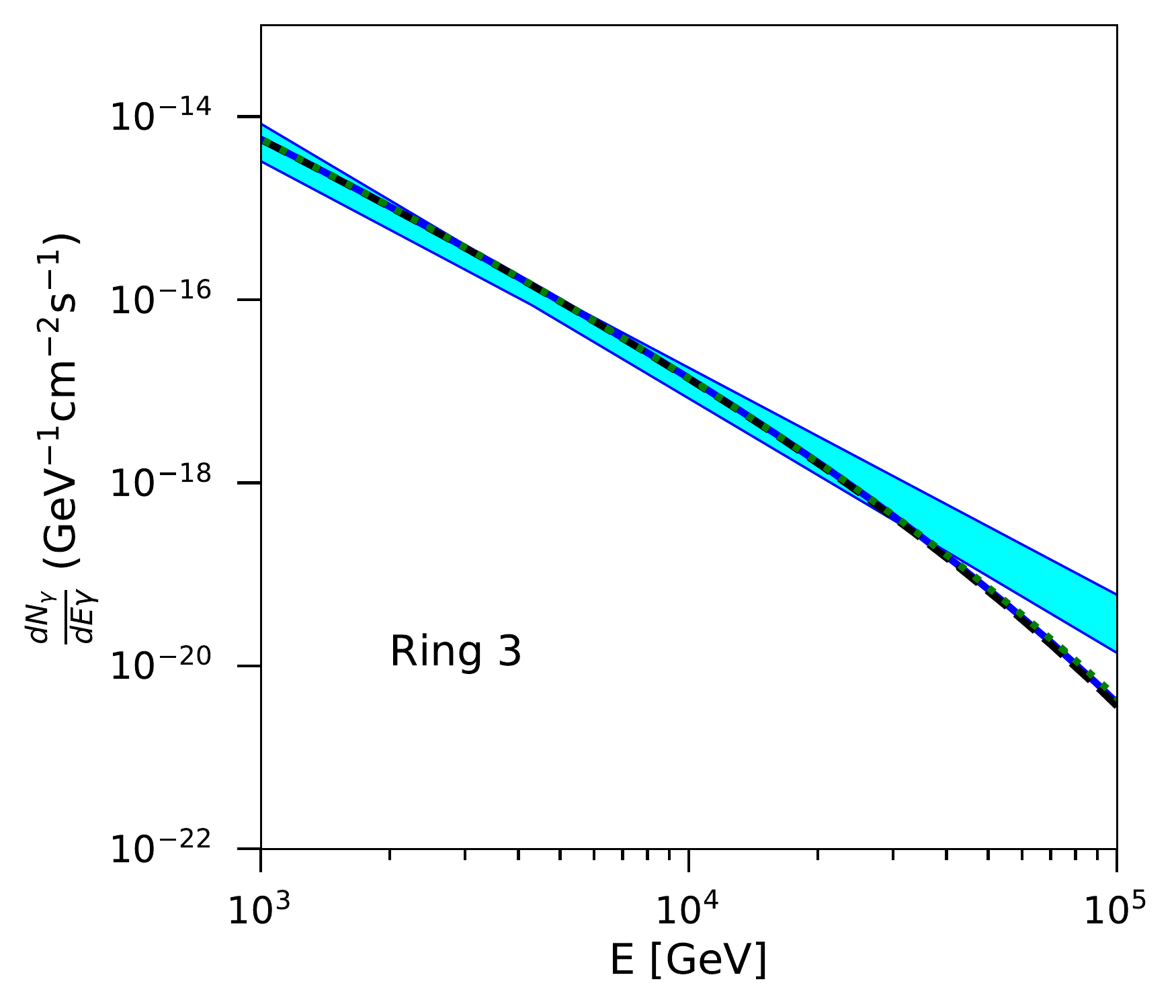}
\includegraphics[width=.45\textwidth]{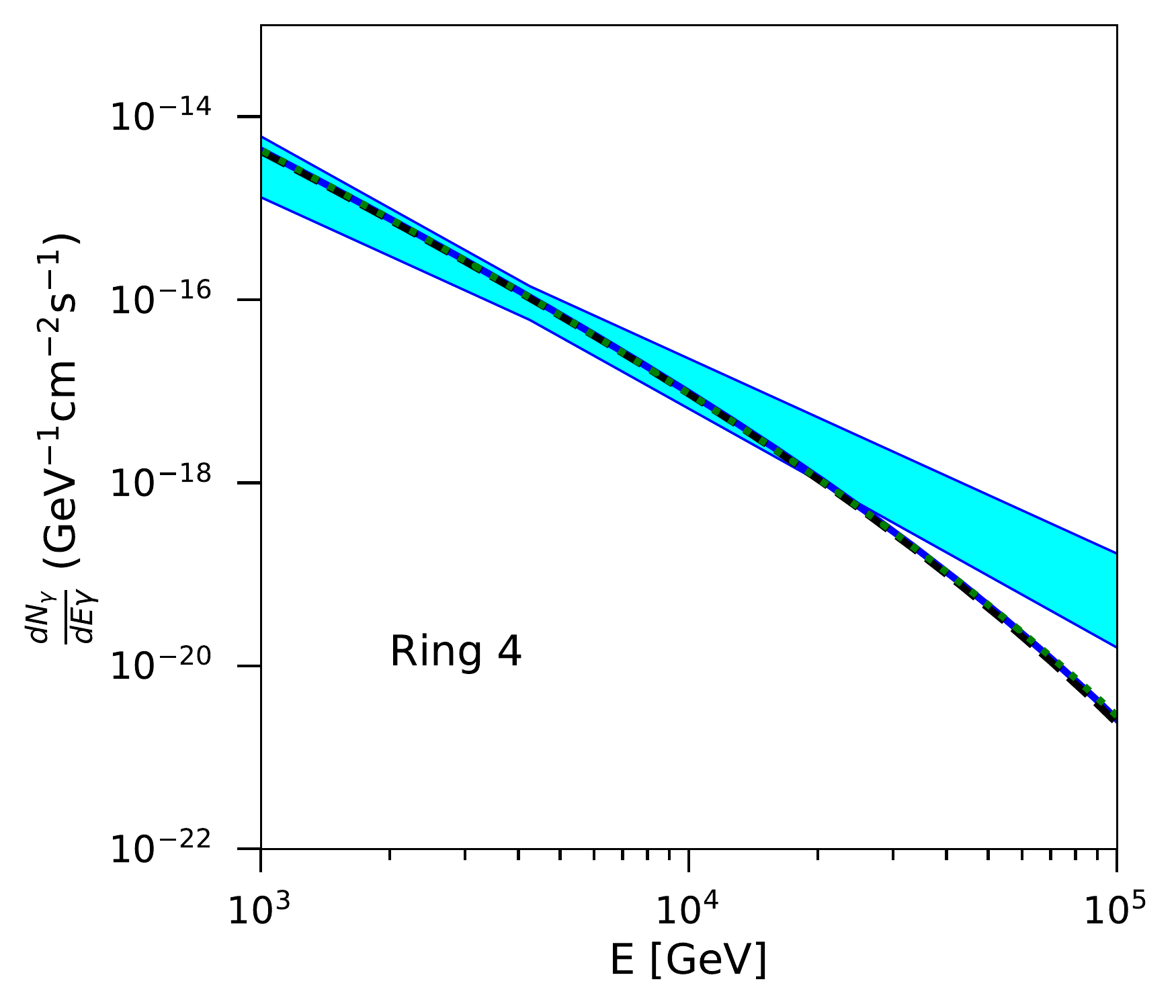}
\caption{Spectrum of gamma ray emission observed by HAWC in four rings around the Cygnus association (shaded areas), compared with the results of our calculations (lines, labelled as in Figure~\ref{fig:GammaFit}). Flux and slope of shaded areas are reported in Table~\ref{tab:HAWC_spectra}.}
\label{fig:MorphSpec}
\end{figure*}

Within the HAWC error bars, the theoretical calculation of the morphology of the gamma ray emission appears to be in excellent agreement with observations. In fact, in addition to the energy integrated information, one can also use some preliminary information on the spectrum in each of the four spatial bins: this information is shown in Figure~\ref{fig:MorphSpec}, where the shaded areas represent the HAWC spectral fit in the different bins and the curves show our results in the same bins, for the same models discussed in Figure~\ref{fig:GammaFit} (the corresponding value of flux and photon spectral index are reported in Table~\ref{tab:HAWC_spectra}, from a private communication with Binita Hona). The agreement between the predicted and the observed gamma ray spectra in the four bins seems evident.

Since the fit to the observations is dominated by the lower energy bins ($\lesssim 10$ TeV) where the error bars are smaller, it is to be expected that some differences with models may appear at the higher energies. In this sense, the future LHAASO measurements of the gamma ray emission from the Cygnus region will play a crucial role in assessing the role of stellar clusters as particle accelerators. 

We notice that a similar analysis of the morphology of the gamma ray emission has also been performed by \cite{Aharonian+2019NatAs} in the region of energies accessible to Fermi-LAT. The analysis suggests that the emission is peaked in the center of the system, rather at odds with the findings at high energies illustrated above. This point will be  extensively discussed in the upcoming article by \cite{Menchiari2023}.

A final comment concerns the impact of the bubble size on the gamma-ray emission. We have seen in \S~\ref{sec:clumps} that, under the reasonable assumption that the target gas is mainly concentrated in clumps with density $n_{\rm cl}\approx 10^{3}\, \rm cm^{-3}$ and size $R_{\rm cl}\approx 1$\,pc, cooling would reduce the bubble size by $\approx 30\%$ with respect to the adiabatic model adopted here. Using the parameters in Table~\ref{tab:Models}, the nominal size of the bubble excavated by the wind is $\sim 100$~pc. Even a reduction of such size by $\sim 30\%$ would still result in a region that is larger than the $\sim 55$ pc from which most gamma ray emission are detected in \cite{Aharonian+2019NatAs} and \cite{2021NatureHAWC}. Hence, since we expect the gamma-ray emission to be roughly $\propto R_{\rm b}$, the small changes can be easily reabsorbed in a small change in either the target gas density or in the CR acceleration efficiency. Also the impact on $p_{\max}$ is marginal and similar values can be obtained by slightly increasing the value of $\eta_{B}$ or requiring a slightly larger diffusion suppression of the downstream diffusion coefficient. 

Very similar considerations can be made in terms of the impact on our results of the uncertainty in the cluster age, which ranges between $2$ and $7$ Myr. According to Eq.~\eqref{eq:Rbubble}, $R_{\rm b}\propto t^{3/5}$, hence changing the age from the 3 Myr to 7 Myr produces a bubble size larger by $\sim 60\%$. However, ages much older than 3 Myr could imply that some SN have exploded, changing the energy input in the bubble system. In such a case our model should be significantly revised to account for the SNR contribution \cite[see, e.g.][]{Vieu+2022}.

\begin{table}
\begin{center}
\begin{tabular}[c]{l c c}  
\toprule \toprule
Ring  &  Flux [cm$^{-2}$ s$^{-1}$ keV$^{-1}$] & $\Gamma$ \\
\midrule
1 & $0.5_{-0.4}^{+1.9} \cdot 10^{-22}$ & $-2.3\pm 0.5$ \vspace{0.2cm} \\
2 & $1.19_{-0.25}^{+0.32} \cdot 10^{-22}$ & $-2.78\pm 0.12$  \vspace{0.2cm} \\
3 & $1.22_{-0.26}^{+0.33} \cdot 10^{-22}$ & $-2.63\pm 0.15$ \vspace{0.2cm} \\ 
4 & $9.4_{-3.2}^{+5} \cdot 10^{-23}$ & $-2.37\pm 0.24$  \vspace{0.1cm} \\ 
\bottomrule %\bottomrule
\end{tabular}
\caption{Best fit parameters for the flux and spectral slope obtained from HAWC data for each ring around the Cygnus OB2 cluster, shown in Figure~\ref{fig:MorphSpec}.}
\label{tab:HAWC_spectra}
\end{center}
\end{table}

\section{Conclusions: What did we learn?}
\label{sec:concl}

Compact star clusters are interesting astrophysical sources for many reasons. Here we investigated their ability to accelerate cosmic rays at the termination shock resulting from the interaction of the collective stellar wind with the surrounding ISM. We also calculated the gamma ray emission produced by the hadronic interactions of the same accelerated particles with the gas embedded in the cavity excavated by the wind and compared the results of our calculations with the recent observations of the Cygnus region carried out by Fermi-LAT and HAWC. 

In a recent article \cite[]{2021MNRASMorlino} we laid down the bases of the theory of particle acceleration at the termination shock, while in the present work we completed such an investigation with the introduction of energy losses, dominated by $pp$ inelastic scattering. Such losses do not play a significant role in shaping the spectrum of accelerated particles at the shock, while they affect the spatial distribution of cosmic rays in the bubble and the associated gamma ray emission if the average density in the wind bubble is of the order of or larger than $\sim 10~\rm cm^{-3}$. 

In our calculation we did not account for the spatial distribution of the target gas in the bubble, retaining the simplest assumption of a constant average gas density. On the other hand, the presence of clumps might affect our conclusions to some extent: first, the role of losses may be more prominent in clumps, where the density is larger, while may be irrelevant for CR propagating in the hot shocked wind. This might turn out to require somewhat larger CR acceleration efficiencies at the termination shock, depending on the filling factor of the clumps. Second, the presence of clumps may affect our considerations on the morphology of the gamma ray emission. Both reasons for concern are certainly to be kept in mind for times in which the information about the spatial location of the clumps and their overdensity will, if any, become known through dedicated observations. At presence, this information is not available and gamma ray observations do not show the level of details that would require more sophisticated modeling. Hence our simple assumption of spatially constant gas density in the bubble.

One of the main reasons for being interested in star clusters is that they have been proposed to be potential PeVatrons. It is very interesting that while at a supernova shock the most severe constraints to particle acceleration come from the confinement of particles upstream of the shock, at the termination shock of a star cluster the particles' return to the shock from the upstream region is guaranteed by the closed spherical geometry (see Figure~\ref{fig:geometry}). In this sense this appears to be an ideal situation to push the particle energy to the limit imposed by their diffusive confinement in the wind region \cite[see also][]{2021MNRASMorlino}. Although this is indeed the case, we show that the spectrum of the accelerated particles does not retain its power law shape up to the nominal value of $p_{\max}$ as due to confinement constraints. While particles diffuse in the upstream (wind) region, they do perceive the spherical geometry as an effective reduced upstream velocity, which in turn means that the effective compression factor gets smaller and the spectrum steeper while moving toward higher energies. For Kolmogorov and Kraichnan turbulence spectra this effect is gradual: it does not result in a cutoff but rather in a rolling steepening of the spectrum so that, although the nominal value of $p_{\max}$ may exceed PeV, the actual spectrum starts becoming steeper much below $p_{\max}$. As discussed by \cite{Morlino+2021}, the only case in which the spectrum of accelerated particles develops an exponential cutoff at $\sim p_{\max}$ is the one that refers to Bohm-like diffusion. Such an energy dependence would be physically justified if the turbulence were self-generated. However, we have shown that for the typical parameters adopted for the Cygnus region, the growth of the non-resonant streaming instability is too slow to result in particle energies exceeding a few tens of TeV.

In conclusion it seems challenging for star clusters with properties similar to the Cygnus region to behave as effective PeVatrons, although we cannot exclude that star clusters requiring more extreme values of the parameters may satisfy the conditions needed for acceleration to the knee region.

In this sense it is of the upmost importance to have at our disposal gamma ray observations that extend to $\gtrsim 100$ TeV, because they provide us with a measurement of the spectrum and spatial distribution of the accelerated particles in the cavity around a star cluster, so that a more serious assessment can be made of their ability to accelerate high energy particles. 

In this article we focused on the Cygnus OB2 association, one of the compact star clusters that have attracted most attention throughout the last decade, mainly because of its high luminosity and wealth of high energy manifestations, recently completed with its detection in the TeV gamma ray band. The Cygnus OB2 association has an estimated luminosity of $\sim (2-5)\times 10^{38}$ erg/s and a wind velocity $v_w\sim 2800-3000$ km/s \citep{Menchiari2023}. If one reads off the maximum energy expected from this object from Figure~\ref{fig:Pmax}, it is easy to see that $p_{\rm \max}\sim$ PeV is possible. The recent gamma ray observation of this source by Fermi-LAT and by HAWC (and more recently by LHAASO) is a great opportunity to test the potential of this accelerator.

We showed that both the spectrum of the gamma ray emission and its morphology in the TeV energy range are well described by adopting reasonable assumptions on the parameters of this system. As mentioned above, the morphology in the GeV energy range, quite different from the one in the TeV band, requires additional consideration and will be discussed elsewhere \citep{Menchiari2023}.

In terms of particle acceleration at the termination shock of this cluster, our conclusion is that although the nominal maximum energy may be in the PeV range, the spectrum of accelerated particles should steepen at energies appreciably lower than PeV. This finding is perfectly in agreement with the observed spectrum and morphology of the TeV gamma ray emission. In terms that are common in describing a particle accelerator, this is not a PeVatron, despite the detection of photons with energy in excess of $\sim 100$ TeV. 

The observed gamma ray emission requires acceleration efficiencies of order of, or below, $\sim 1\%$, about $3-10$ smaller than the efficiency typically associated to SNR shocks. It is however worth stressing that this estimate was derived by following the same procedure as in HAWC data analysis: the emission region is assumed to be Gaussian and the total gamma ray flux is affected by this assumption. In particular, we showed that this procedure leads to overestimating the gamma ray flux (namely underestimating the CR acceleration efficiency) and to infer an injection spectrum somewhat steeper than the actual one. If, on the other hand, this low efficiency were to be confirmed as a physical effect, it might either reflect the smaller shock velocity in the case of compact star clusters or  suggest that the mean density in the bubble is somewhat lower than adopted here. This last possibility would also mean that the values adopted in earlier analyses should be reconsidered as well.

If stellar clusters were the main sources of Galactic CRs, the low efficiency estimated here would require $\sim 10^4$ star clusters similar to Cygnus OB2 to have been active during the confinement time of CRs in the Galaxy. Clearly the constraint weakens if they only provide a fraction of Galactic CRs. 

The upcoming more detailed observations of the Cygnus region and of other star clusters, possibly extending the measurements to higher energies, will be of great importance to build a stronger case in favor or against compact star clusters as contributors to Galactic CRs. 

\section*{Acknowledgements}
We thank the referee for the comments that helped us improving our manuscript. The authors acknowledge fruitful discussion with S. Menchiari, E. Amato, N. Bucciantini, C. Evoli and O. Fornieri. GM is partially supported by the INAF Theory Grant 2022 {\it ‘‘Star Clusters As Cosmic Ray Factories''}.

\section*{Data availability}
All data used in this article can be retrieved in the cited publications with the exception of Figure~\ref{fig:MorphSpec} whose data come from a private communication and are reported in Table~\ref{tab:HAWC_spectra}.

\bibliographystyle{mnras}
\bibliography{biblio} % if your bibtex file is called example.bib

\begin{thebibliography}{}
\makeatletter
\relax
\def\mn@urlcharsother{\let\do\@makeother \do\$\do\&\do\#\do\^\do\_\do\%\do\~}
\def\mn@doi{\begingroup\mn@urlcharsother \@ifnextchar [ {\mn@doi@}
  {\mn@doi@[]}}
\def\mn@doi@[#1]#2{\def\@tempa{#1}\ifx\@tempa\@empty \href
  {http://dx.doi.org/#2} {doi:#2}\else \href {http://dx.doi.org/#2} {#1}\fi
  \endgroup}
\def\mn@eprint#1#2{\mn@eprint@#1:#2::\@nil}
\def\mn@eprint@arXiv#1{\href {http://arxiv.org/abs/#1} {{\tt arXiv:#1}}}
\def\mn@eprint@dblp#1{\href {http://dblp.uni-trier.de/rec/bibtex/#1.xml}
  {dblp:#1}}
\def\mn@eprint@#1:#2:#3:#4\@nil{\def\@tempa {#1}\def\@tempb {#2}\def\@tempc
  {#3}\ifx \@tempc \@empty \let \@tempc \@tempb \let \@tempb \@tempa \fi \ifx
  \@tempb \@empty \def\@tempb {arXiv}\fi \@ifundefined
  {mn@eprint@\@tempb}{\@tempb:\@tempc}{\expandafter \expandafter \csname
  mn@eprint@\@tempb\endcsname \expandafter{\@tempc}}}

\bibitem[\protect\citeauthoryear{{Abeysekara} et~al.,}{{Abeysekara}
  et~al.}{2021}]{2021NatureHAWC}
{Abeysekara} A.~U.,  et~al., 2021, \mn@doi [Nature Astronomy]
  {10.1038/s41550-021-01318-y}, \href
  {https://ui.adsabs.harvard.edu/abs/2021NatAs...5..465A} {5, 465}

\bibitem[\protect\citeauthoryear{{Abramowski} et~al.,}{{Abramowski}
  et~al.}{2012}]{Abramowski_Wd1:2012}
{Abramowski} A.,  et~al., 2012, \mn@doi [\aap] {10.1051/0004-6361/201117928},
  \href {https://ui.adsabs.harvard.edu/abs/2012A&A...537A.114A} {537, A114}

\bibitem[\protect\citeauthoryear{Ackermann \& et al.}{Ackermann \&
  et~al.}{2011}]{Ackermann:2011p3159}
Ackermann M.,  et al. 2011, \mn@doi [Science] {10.1126/science.1210311}, 334,
  1103

\bibitem[\protect\citeauthoryear{Ackermann et~al.,}{Ackermann
  et~al.}{2011}]{ackermann-Cygnus2011}
Ackermann M.,  et~al., 2011, \mn@doi [Science] {10.1126/science.1210311}, 334,
  1103

\bibitem[\protect\citeauthoryear{{Aharonian}, {Yang}  \& {de O{\~n}a
  Wilhelmi}}{{Aharonian} et~al.}{2019}]{Aharonian+2019NatAs}
{Aharonian} F.,  {Yang} R.,   {de O{\~n}a Wilhelmi} E.,  2019, \mn@doi [Nature
  Astronomy] {10.1038/s41550-019-0724-0}, \href
  {https://ui.adsabs.harvard.edu/abs/2019NatAs...3..561A} {3, 561}

\bibitem[\protect\citeauthoryear{{Albert} et~al.,}{{Albert}
  et~al.}{2021}]{HAWC:2021}
{Albert} A.,  et~al., 2021, \mn@doi [\apjl] {10.3847/2041-8213/abd77b}, \href
  {https://ui.adsabs.harvard.edu/abs/2021ApJ...907L..30A} {907, L30}

\bibitem[\protect\citeauthoryear{{Badmaev}, {Bykov}  \& {Kalyashova}}{{Badmaev}
  et~al.}{2022}]{Badmaev+2022}
{Badmaev} D.~V.,  {Bykov} A.~M.,   {Kalyashova} M.~E.,  2022, \mn@doi [\mnras]
  {10.1093/mnras/stac2738}, \href
  {https://ui.adsabs.harvard.edu/abs/2022MNRAS.517.2818B} {517, 2818}

\bibitem[\protect\citeauthoryear{Bell}{Bell}{1978}]{Bell:1978p1342}
Bell A.~R.,  1978, Royal Astronomical Society, 182, 443

\bibitem[\protect\citeauthoryear{{Bell}}{{Bell}}{2004a}]{bell2004}
{Bell} A.~R.,  2004a, \mn@doi [MNRAS] {10.1111/j.1365-2966.2004.08097.x}, \href
  {http://adsabs.harvard.edu/abs/2004MNRAS.353..550B} {353, 550}

\bibitem[\protect\citeauthoryear{Bell}{Bell}{2004b}]{Bell:2004p737}
Bell A.~R.,  2004b, \mn@doi [Monthly Notices of the Royal Astronomical Society]
  {10.1111/j.1365-2966.2004.08097.x}, 353, 550

\bibitem[\protect\citeauthoryear{{Binns} et~al.,}{{Binns}
  et~al.}{2006}]{Binns+:2006}
{Binns} W.~R.,  et~al., 2006, \mn@doi [\nar] {10.1016/j.newar.2006.06.058},
  \href {https://ui.adsabs.harvard.edu/abs/2006NewAR..50..516B} {50, 516}

\bibitem[\protect\citeauthoryear{{Blandford} \& {Eichler}}{{Blandford} \&
  {Eichler}}{1987}]{blandford}
{Blandford} R.,  {Eichler} D.,  1987, \mn@doi [\physrep]
  {10.1016/0370-1573(87)90134-7}, \href
  {https://ui.adsabs.harvard.edu/abs/1987PhR...154....1B} {154, 1}

\bibitem[\protect\citeauthoryear{{Bykov} \& {Kalyashova}}{{Bykov} \&
  {Kalyashova}}{2022}]{Bykov-Kalyashova:2022}
{Bykov} A.~M.,  {Kalyashova} M.~E.,  2022, \mn@doi [Advances in Space Research]
  {10.1016/j.asr.2022.01.029}, \href
  {https://ui.adsabs.harvard.edu/abs/2022AdSpR..70.2685B} {70, 2685}

\bibitem[\protect\citeauthoryear{{Bykov}, {Marcowith}, {Amato}, {Kalyashova},
  {Kruijssen}  \& {Waxman}}{{Bykov} et~al.}{2020}]{Bykov+2020}
{Bykov} A.~M.,  {Marcowith} A.,  {Amato} E.,  {Kalyashova} M.~E.,  {Kruijssen}
  J.~M.~D.,   {Waxman} E.,  2020, \mn@doi [\ssr] {10.1007/s11214-020-00663-0},
  \href {https://ui.adsabs.harvard.edu/abs/2020SSRv..216...42B} {216, 42}

\bibitem[\protect\citeauthoryear{{Cao} et~al.,}{{Cao}
  et~al.}{2021a}]{LHAASO-2021Nature}
{Cao} Z.,  et~al., 2021a, \mn@doi [\nat] {10.1038/s41586-021-03498-z}, \href
  {https://ui.adsabs.harvard.edu/abs/2021Natur.594...33C} {594, 33}

\bibitem[\protect\citeauthoryear{{Cao} et~al.,}{{Cao}
  et~al.}{2021b}]{2021Natur.594...33C}
{Cao} Z.,  et~al., 2021b, \mn@doi [\nat] {10.1038/s41586-021-03498-z}, \href
  {https://ui.adsabs.harvard.edu/abs/2021Natur.594...33C} {594, 33}

\bibitem[\protect\citeauthoryear{{Cesarsky} \& {Montmerle}}{{Cesarsky} \&
  {Montmerle}}{1983}]{Cesarsky-Montmerle:1983}
{Cesarsky} C.~J.,  {Montmerle} T.,  1983, \mn@doi [Space Sc. Rev.]
  {10.1007/BF00167503}, \href
  {https://ui.adsabs.harvard.edu/abs/1983SSRv...36..173C} {36, 173}

\bibitem[\protect\citeauthoryear{{Cowie} \& {McKee}}{{Cowie} \&
  {McKee}}{1977}]{Cowie-McKee:1977}
{Cowie} L.~L.,  {McKee} C.~F.,  1977, \mn@doi [\apj] {10.1086/154911}, \href
  {https://ui.adsabs.harvard.edu/abs/1977ApJ...211..135C} {211, 135}

\bibitem[\protect\citeauthoryear{{Cristofari}, {Blasi}  \&
  {Amato}}{{Cristofari} et~al.}{2020}]{pierre}
{Cristofari} P.,  {Blasi} P.,   {Amato} E.,  2020, \mn@doi [Astroparticle
  Physics] {10.1016/j.astropartphys.2020.102492}, \href
  {https://ui.adsabs.harvard.edu/abs/2020APh...12302492C} {123, 102492}

\bibitem[\protect\citeauthoryear{{Cristofari}, {Blasi}  \&
  {Caprioli}}{{Cristofari} et~al.}{2021}]{pierre2021}
{Cristofari} P.,  {Blasi} P.,   {Caprioli} D.,  2021, \mn@doi [\aap]
  {10.1051/0004-6361/202140448}, \href
  {https://ui.adsabs.harvard.edu/abs/2021A&A...650A..62C} {650, A62}

\bibitem[\protect\citeauthoryear{{Draine}}{{Draine}}{2011}]{Draine_book:2011}
{Draine} B.~T.,  2011, {Physics of the Interstellar and Intergalactic Medium}

\bibitem[\protect\citeauthoryear{{Dundovic}, {Pezzi}, {Blasi}, {Evoli}  \&
  {Matthaeus}}{{Dundovic} et~al.}{2020}]{Dundovic2020}
{Dundovic} A.,  {Pezzi} O.,  {Blasi} P.,  {Evoli} C.,   {Matthaeus} W.~H.,
  2020, \mn@doi [\prd] {10.1103/PhysRevD.102.103016}, \href
  {https://ui.adsabs.harvard.edu/abs/2020PhRvD.102j3016D} {102, 103016}

\bibitem[\protect\citeauthoryear{Giacalone \& Jokipii}{Giacalone \&
  Jokipii}{2007}]{Giacalone:2007p962}
Giacalone J.,  Jokipii J.~R.,  2007, \mn@doi [The Astrophysical Journal]
  {10.1086/519994}, 663, L41

\bibitem[\protect\citeauthoryear{{Gupta}, {Nath}, {Sharma}  \&
  {Shchekinov}}{{Gupta} et~al.}{2016}]{Gupta+2016}
{Gupta} S.,  {Nath} B.~B.,  {Sharma} P.,   {Shchekinov} Y.,  2016, \mn@doi
  [\mnras] {10.1093/mnras/stw1920}, \href
  {https://ui.adsabs.harvard.edu/abs/2016MNRAS.462.4532G} {462, 4532}

\bibitem[\protect\citeauthoryear{{Gupta}, {Nath}, {Sharma}  \&
  {Eichler}}{{Gupta} et~al.}{2018}]{Gupta+2018}
{Gupta} S.,  {Nath} B.~B.,  {Sharma} P.,   {Eichler} D.,  2018, \mn@doi
  [\mnras] {10.1093/mnras/stx2427}, \href
  {https://ui.adsabs.harvard.edu/abs/2018MNRAS.473.1537G} {473, 1537}

\bibitem[\protect\citeauthoryear{{Gupta}, {Nath}, {Sharma}  \&
  {Eichler}}{{Gupta} et~al.}{2020}]{Gupta+2020}
{Gupta} S.,  {Nath} B.~B.,  {Sharma} P.,   {Eichler} D.,  2020, \mn@doi
  [\mnras] {10.1093/mnras/staa286}, \href
  {https://ui.adsabs.harvard.edu/abs/2020MNRAS.493.3159G} {493, 3159}

\bibitem[\protect\citeauthoryear{{H.~E.~S.~S. Collaboration}
  et~al.,}{{H.~E.~S.~S. Collaboration} et~al.}{2015}]{HESS-30Dor:2015}
{H.~E.~S.~S. Collaboration} et~al., 2015, \mn@doi [Science]
  {10.1126/science.1261313}, \href
  {https://ui.adsabs.harvard.edu/abs/2015Sci...347..406H} {347, 406}

\bibitem[\protect\citeauthoryear{{Kalyashova} \& {Bykov}}{{Kalyashova} \&
  {Bykov}}{2021}]{Kalyashova-Bykov:2021}
{Kalyashova} M.~E.,  {Bykov} A.~M.,  2021, in Journal of Physics Conference
  Series. p. 012008, \mn@doi{10.1088/1742-6596/2103/1/012008}

\bibitem[\protect\citeauthoryear{{Kelner}, {Aharonian}  \& {Bugayov}}{{Kelner}
  et~al.}{2006}]{kelner2006}
{Kelner} S.~R.,  {Aharonian} F.~A.,   {Bugayov} V.~V.,  2006, \mn@doi [\prd]
  {10.1103/PhysRevD.74.034018}, \href
  {https://ui.adsabs.harvard.edu/abs/2006PhRvD..74c4018K} {74, 034018}

\bibitem[\protect\citeauthoryear{{Koo} \& {McKee}}{{Koo} \&
  {McKee}}{1992a}]{Koo-McKee:1992a}
{Koo} B.-C.,  {McKee} C.~F.,  1992a, \mn@doi [\apj] {10.1086/171132}, \href
  {https://ui.adsabs.harvard.edu/abs/1992ApJ...388...93K} {388, 93}

\bibitem[\protect\citeauthoryear{{Koo} \& {McKee}}{{Koo} \&
  {McKee}}{1992b}]{Koo-McKee:1992b}
{Koo} B.-C.,  {McKee} C.~F.,  1992b, \mn@doi [\apj] {10.1086/171133}, \href
  {https://ui.adsabs.harvard.edu/abs/1992ApJ...388..103K} {388, 103}

\bibitem[\protect\citeauthoryear{{Krakau} \& {Schlickeiser}}{{Krakau} \&
  {Schlickeiser}}{2015}]{krakau2015}
{Krakau} S.,  {Schlickeiser} R.,  2015, \mn@doi [\apj]
  {10.1088/0004-637X/802/2/114}, \href
  {https://ui.adsabs.harvard.edu/abs/2015ApJ...802..114K} {802, 114}

\bibitem[\protect\citeauthoryear{{Lagage} \& {Cesarsky}}{{Lagage} \&
  {Cesarsky}}{1983a}]{Lagage1}
{Lagage} P.~O.,  {Cesarsky} C.~J.,  1983a, A\&A, \href
  {http://adsabs.harvard.edu/abs/1983A%26A...118..223L} {118, 223}

\bibitem[\protect\citeauthoryear{{Lagage} \& {Cesarsky}}{{Lagage} \&
  {Cesarsky}}{1983b}]{Lagage2}
{Lagage} P.~O.,  {Cesarsky} C.~J.,  1983b, A\&A, \href
  {http://adsabs.harvard.edu/abs/1983A%26A...125..249L} {125, 249}

\bibitem[\protect\citeauthoryear{{Menchiari}, {Morlino}, {Amato}  \&
  {Bucciantini}}{{Menchiari} et~al.}{2023}]{Menchiari2023}
{Menchiari} S.,  {Morlino} G.,  {Amato} E.,   {Bucciantini} N.,  2023, in
  preparation

\bibitem[\protect\citeauthoryear{{Morlino}, {Blasi}, {Peretti}  \&
  {Cristofari}}{{Morlino} et~al.}{2021a}]{Morlino+2021}
{Morlino} G.,  {Blasi} P.,  {Peretti} E.,   {Cristofari} P.,  2021a, \mn@doi
  [\mnras] {10.1093/mnras/stab690}, \href
  {https://ui.adsabs.harvard.edu/abs/2021MNRAS.tmp..749M} {}

\bibitem[\protect\citeauthoryear{{Morlino}, {Blasi}, {Peretti}  \&
  {Cristofari}}{{Morlino} et~al.}{2021b}]{2021MNRASMorlino}
{Morlino} G.,  {Blasi} P.,  {Peretti} E.,   {Cristofari} P.,  2021b, \mn@doi
  [\mnras] {10.1093/mnras/stab690}, \href
  {https://ui.adsabs.harvard.edu/abs/2021MNRAS.504.6096M} {504, 6096}

\bibitem[\protect\citeauthoryear{{Prantzos}}{{Prantzos}}{2012}]{Prantzos2012}
{Prantzos} N.,  2012, \mn@doi [\aap] {10.1051/0004-6361/201117448}, \href
  {https://ui.adsabs.harvard.edu/abs/2012A&A...538A..80P} {538, A80}

\bibitem[\protect\citeauthoryear{{Saha}, {Dom{\'\i}nguez}, {Tibaldo},
  {Marchesi}, {Ajello}, {Lemoine-Goumard}  \& {L{\'o}pez}}{{Saha}
  et~al.}{2020}]{Saha_NGC3603:2020}
{Saha} L.,  {Dom{\'\i}nguez} A.,  {Tibaldo} L.,  {Marchesi} S.,  {Ajello} M.,
  {Lemoine-Goumard} M.,   {L{\'o}pez} M.,  2020, \mn@doi [\apj]
  {10.3847/1538-4357/ab9ac2}, \href
  {https://ui.adsabs.harvard.edu/abs/2020ApJ...897..131S} {897, 131}

\bibitem[\protect\citeauthoryear{{Schure} \& {Bell}}{{Schure} \&
  {Bell}}{2013}]{Schure-Bell:2013}
{Schure} K.~M.,  {Bell} A.~R.,  2013, \mn@doi [\mnras] {10.1093/mnras/stt1371},
  \href {https://ui.adsabs.harvard.edu/abs/2013MNRAS.435.1174S} {435, 1174}

\bibitem[\protect\citeauthoryear{Spitzer}{Spitzer}{1962}]{spitzerBook}
Spitzer L.,  1962, Physics of fully ionized gases.
John Wiley \& Sons

\bibitem[\protect\citeauthoryear{{Subedi} et~al.,}{{Subedi}
  et~al.}{2017}]{Subedi2017}
{Subedi} P.,  et~al., 2017, \mn@doi [\apj] {10.3847/1538-4357/aa603a}, \href
  {https://ui.adsabs.harvard.edu/abs/2017ApJ...837..140S} {837, 140}

\bibitem[\protect\citeauthoryear{{Sun}, {Yang}, {Liang}, {Peng}, {Zhang},
  {Wang}  \& {Aharonian}}{{Sun} et~al.}{2020}]{Sun_W40:2020}
{Sun} X.-N.,  {Yang} R.-Z.,  {Liang} Y.-F.,  {Peng} F.-K.,  {Zhang} H.-M.,
  {Wang} X.-Y.,   {Aharonian} F.,  2020, \mn@doi [\aap]
  {10.1051/0004-6361/202037580}, \href
  {https://ui.adsabs.harvard.edu/abs/2020A&A...639A..80S} {639, A80}

\bibitem[\protect\citeauthoryear{{Tatischeff}, {Raymond}, {Duprat}, {Gabici}
  \& {Recchia}}{{Tatischeff} et~al.}{2021}]{Tatischeff+2021}
{Tatischeff} V.,  {Raymond} J.~C.,  {Duprat} J.,  {Gabici} S.,   {Recchia} S.,
  2021, \mn@doi [\mnras] {10.1093/mnras/stab2533}, \href
  {https://ui.adsabs.harvard.edu/abs/2021MNRAS.508.1321T} {508, 1321}

\bibitem[\protect\citeauthoryear{{Vieu}, {Gabici}, {Tatischeff}  \&
  {Ravikularaman}}{{Vieu} et~al.}{2022}]{Vieu+2022}
{Vieu} T.,  {Gabici} S.,  {Tatischeff} V.,   {Ravikularaman} S.,  2022, \mn@doi
  [\mnras] {10.1093/mnras/stac543}, \href
  {https://ui.adsabs.harvard.edu/abs/2022MNRAS.512.1275V} {512, 1275}

\bibitem[\protect\citeauthoryear{{Weaver}, {McCray}, {Castor}, {Shapiro}  \&
  {Moore}}{{Weaver} et~al.}{1977}]{Weaver+1977}
{Weaver} R.,  {McCray} R.,  {Castor} J.,  {Shapiro} P.,   {Moore} R.,  1977,
  \mn@doi [\apj] {10.1086/155692}, \href
  {https://ui.adsabs.harvard.edu/abs/1977ApJ...218..377W} {218, 377}

\bibitem[\protect\citeauthoryear{{Webb}, {Axford}  \& {Forman}}{{Webb}
  et~al.}{1985}]{webb1985}
{Webb} G.~M.,  {Axford} W.~I.,   {Forman} M.~A.,  1985, \mn@doi [\apj]
  {10.1086/163652}, \href
  {https://ui.adsabs.harvard.edu/abs/1985ApJ...298..684W} {298, 684}

\bibitem[\protect\citeauthoryear{{Yang}, {de O{\~n}a Wilhelmi}  \&
  {Aharonian}}{{Yang} et~al.}{2018}]{Yang+2018}
{Yang} R.-z.,  {de O{\~n}a Wilhelmi} E.,   {Aharonian} F.,  2018, \mn@doi
  [\aap] {10.1051/0004-6361/201732045}, \href
  {https://ui.adsabs.harvard.edu/abs/2018A&A...611A..77Y} {611, A77}

\makeatother
\end{thebibliography}

% Alternatively you could enter them by hand, like this:
% This method is tedious and prone to error if you have lots of references
%\begin{thebibliography}{99}
%\bibitem[\protect\citeauthoryear{Author}{2012}]{Author2012}
%Author A.~N., 2013, Journal of Improbable Astronomy, 1, 1
%\bibitem[\protect\citeauthoryear{Others}{2013}]{Others2013}
%Others S., 2012, Journal of Interesting Stuff, 17, 198
%\end{thebibliography}

%%%%%%%%%%%%%%%%%%%%%%%%%%%%%%%%%%%%%%%%%%%%%%%%%%

%%%%%%%%%%%%%%%%% APPENDICES %%%%%%%%%%%%%%%%%%%%%

%\appendix
%\section{Some extra material}

%%%%%%%%%%%%%%%%%%%%%%%%%%%%%%%%%%%%%%%%%%%%%%%%%%

\appendix 
\section{Numerical scheme for the solution of the transport equation}
\label{app:A}

As discussed in \S \ref{sec:theory}, the integration of the transport equation around the location of the shock leads to a solution for $N_0(E)$ in the form given in Eq. \ref{eq:N0}. This expression depends explicitly upon the derivatives of the distribution function immediately upstream and downstream of the shock. These two quantities are calculated at each iteration from the solution for a given $N_0$. The iterative procedure starts with $N_0$ in the form of a power law with the canonical slope associated to the shock compression factor. 

The numerical solution of the transport equation at each iteration reduces to solving for the spatial distribution functions upstream and downstream respectively. We introduce the following quantities:
\begin{eqnarray}
    \alpha(r)=r^2 u, ~~~~~~~ \alpha'(r)=\frac{d\alpha}{dr}\\
    \beta(E,r)=r^2 D(E,r),~~~~~~~\beta'(E,r)=\frac{d\beta}{dr},       
\end{eqnarray}
and we use the variable $t=\ln E$, so that the transport equation reads:
\begin{equation}
    -\beta \frac{\partial^2 N}{\partial^2 r}+\frac{\partial N}{\partial r} \left(\alpha-\beta'\right)+N\left( \frac{\alpha'}{3}+r^2 b'\right)-\frac{\partial N}{\partial t} \left( \frac{\alpha'}{3} - \frac{r^2 b}{E} \right)=0.
\end{equation}
In order to describe properly the transport of low energy and high energy particles we adopt a log scale in the radius, namely we introduce a variable $\eta = \ln \left| 1 - \frac{r}{R_s}\right|$. In terms of these new variables the equation becomes:
\begin{equation}
    C_1 \frac{\partial^2 N}{\partial \eta^2} + C_2 \frac{\partial N}{\partial \eta} + C_3 N -C_4 \frac{\partial N}{\partial t} = 0,
\end{equation} 
where the coefficients are defined as follows:
\begin{eqnarray}
    C_1=-\beta \frac{e^{-2\eta}}{R_s^2},\\
    C_2=\beta \frac{e^{-2\eta}}{R_s^2} \mp (\alpha-\beta')\frac{e^{-\eta}}{R_s},\\
    C_3=\frac{2}{3}\alpha' + r^2 b'\\
    C_4=\frac{1}{3}\alpha'-\frac{r^2 b}{E},
\end{eqnarray}
where the upper (lower) sign in $C_2$ refers to the upstream (downstream) region. Notice that downstream $u\propto r^{-2}$, hence $\alpha'=0$ there. The relation between $\eta$ and $r$ is such that $r=R_s \left( 1\mp e^\eta \right)$, where again the upper (lower) sign applies to upstream (downstream). 

The discretization of the transport equation leads to a set of tridiagonal equations in the form:
\begin{eqnarray}
    \left[ C_1 - \frac{C_2 \Delta \eta}{2}\right] N_{j,k-1} + \nonumber\\
    \left[ -2 C_1 + C_3 \Delta \eta^2 + C_4 \frac{\Delta \eta^2}{\Delta t}\right] N_{j,k} + \nonumber \\
    \left[ C_1 + \frac{C_2 \Delta \eta}{2}\right]N_{j,k+1}=C_4 N_{j+1,k} \frac{\Delta \eta^2}{\Delta t}.
    \label{eq:discrete}
\end{eqnarray}
Here $j$ and $k$ denote respectively the momentum and space indexes, and $\Delta \eta$ and $\Delta t$ refer to the spacing in the $\eta$ and $t$ grids respectively. 

The set of algebraic equations is solved starting from the condition that at the highest momentum in the grid (chosen to be at momentum much larger than the physical maximum momentum in the problem at hand) $N=0$ in all points of the spatial grid. The boundary conditions in space are such that $N(r=R_b,t)=0$ (free escape from the bubble) and vanishing flux at $r=0$ (see Eq. \ref{eq:flux}). The normalization of the distribution function is calculated {\it a posteriori} by requiring that a given fraction $\xi_{CR}$ of the upstream kinetic pressure is converted to accelerated particles. 

The logarithmic nature of the spatial scale is such that the shock location, $r=R_s$, cannot be formally part of the grid. However, the first point in the upstream and downstream grid can be chosen as close as needed to the shock location. One can easily check the extremely weak dependence of the results upon the proximity of the first point to the shock location, provided the distance is much smaller than the diffusion length of the lowest energy particles in the momentum grid. 

The set of Eqs. \ref{eq:discrete} is solved numerically upstream and downstream for a given $N_0$. The derivatives upstream (${\cal D}_1$) and downstream (${\cal D}_2$) are calculated numerically and used to update the function $N_0(E)$ using Eq. \ref{eq:N0}. The procedure is repeated until convergence on $N_0$ and the spatial distribution is reached (typically less than 10 iterations are necessary).

% Don't change these lines
\bsp	% typesetting comment
\label{lastpage}
\end{document}